\documentclass{aa}  

\usepackage{graphicx}
\usepackage{txfonts}
\usepackage{color}
\usepackage{xcolor}
\definecolor{xlinkcolor}{cmyk}{1,1,0,0}
\usepackage[bookmarks=true,         % show bookmarks bar?
bookmarksopen=true,
pdfnewwindow=true,      % links in new window
colorlinks=true,    % false: boxed links; true: colored links
linkcolor=xlinkcolor,     % color of internal links
citecolor=xlinkcolor,     % color of links to bibliography
filecolor=xlinkcolor,  % color of file links
urlcolor=xlinkcolor,      % color of external links
final=true]{hyperref} 
\hypersetup{pdfstartview=Fit}
\usepackage{bookmark}
\usepackage[normalem]{ulem}
\usepackage{subcaption}
\usepackage{physics}
\usepackage{subdepth}

\defcitealias{dziembowski1992}{DG92}

\newcommand{\di}[0]{\mathrm{i}}
\newcommand{\rr}[0]{\mathrm{r}}
\newcommand{\hh}[0]{\mathrm{h}}

\newcommand{\orcit}[1]{\protect\href{https://orcid.org/#1}{\protect\includegraphics[width=3mm]{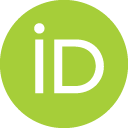}}}

\begin{document}

	\title{Impact of near-degeneracy effects on linear rotational inversions for red-giant stars}
	
	\author{F. Ahlborn \inst{1\,\orcit{0000-0003-0343-6945}}
		\and J. M. Joel Ong\inst{2,3\,\orcit{0000-0001-7664-648X}}
		\and J. Van Beeck\inst{1\,\orcit{0000-0002-5082-3887}}
		\and E. P. Bellinger \inst{4,5\,\orcit{0000-0003-4456-4863}}
		\and S. Hekker \inst{1,6\,\orcit{0000-0002-1463-726X}}
		\and S. Basu \inst{4\,\orcit{0000-0002-6163-3472}}}
	\institute{Heidelberger Institut f\"ur Theoretische Studien, Schloss-Wolfsbrunnenweg 35, 69118 Heidelberg, Germany\\
		\email{felix.ahlborn@h-its.org}
		\and
		Institute for Astronomy, University of Hawai’i, 2680 Woodlawn Drive, Honolulu, HI 96822, USA
		\and
		NASA Hubble Fellow
		\and
		Department of Astronomy, Yale University, New Haven, CT 06520, USA
		\and
		Max-Planck-Institut f\"ur Astrophysik, Karl-Schwarzschild-Stra{\ss}e 1, 85748 Garching, Germany
		\and
		Center for Astronomy (ZAH/LSW), Heidelberg University, Königstuhl 12, 69117 Heidelberg, Germany
	}

	\date{Received; accepted}
	
	\abstract
	{Accurate estimates of internal red-giant rotation rates are a crucial ingredient for constraining and improving current models of stellar rotation. Asteroseismic rotational inversions are a method to estimate these internal rotation rates.} %leave it empty if necessary  
	{In this work, we focus on the observed differences in the rotationally-induced frequency shifts between prograde and retrograde modes, which were ignored in previous works when estimating internal rotation rates of red giants using inversions. We systematically study the limits of applicability of linear rotational inversions as a function of the evolution on the red-giant branch and the underlying rotation rates.}
	{We solve for the oscillation mode frequencies in the presence of rotation in the lowest-order perturbative approach. This enables a description of the differences between prograde and retrograde modes through the coupling of multiple mixed modes, also known as near-degeneracy effects. We compute synthetic rotational splittings taking these near-degeneracy effects into account. We use red-giant models with one solar mass, a large frequency separation between 16 and 9~$\mu$Hz and core rotation rates between 500 and 1500~nHz covering the regime of observed parameters of {\it Kepler} red-giant stars. Finally, we use these synthetic data to quantify the systematic errors of internal rotation rates estimated by means of rotational inversions in the presence of near-degeneracy effects.}
	{ We show that the systematic errors in the estimated rotation rates introduced by near-degeneracy effects surpass observational uncertainties for more evolved and faster rotating stars. For a core rotation rate of 500~nHz linear inversions remain applicable over the range of models considered here, while for a core rotation rate of 1000~nHz systematic errors start to become significant below a large frequency separation of 13~$\mu$Hz.}
	{The estimated rotation rates of some of the previously analysed red giants suffer from significant systematic errors that have not been taken into account yet. Notwithstanding, reliable analyses with existing inversion methods are feasible for a number of red giants, and we expect there to be unexplored targets within the parameter ranges determined here. Finally, utilizing the observational potential of the near-degeneracy effects constitutes an important step toward obtaining more accurate estimates of internal red-giant rotation rates.}
	
	\keywords{asteroseismology -- stars: rotation -- stars: oscillations -- stars: interiors}
	
	\maketitle
	%
	%-------------------------------------------------------------------
	
	\section{Introduction}\label{secintro}
	We can gain insight into the interiors of stars through their observed oscillations using a collection of methods known as `inversions' which probe their internal structure, magnetic field strength, and rotation. These inversion techniques seek to measure a perturbation of the stellar structure (e.g. due to rotation) compared to some known fiducial state. To use these inversion methods, a set of assumptions need to be fulfilled. Most notably, the use of linear inversions requires a linear relation to exist between the perturbation to the frequencies, and the size of the perturbation. For rotational inversions of red-giant stars, however, we may enter a regime in which this assumption of linearity is no longer valid \citep{deheuvels2017,li2024}.
	
	Rotation is an important, yet poorly understood, property of stellar interiors. Chemical mixing induced by rotation may significantly change the distribution of the chemical elements inside the star, such as by increasing the supply of elements available for nuclear burning in the core \cite[e.g.][]{eggenberger2010}. This may also alter the surface abundances measured through classical spectroscopy. Rotation also perturbs the structure of the star directly. For example, in the case of fast rotation it perturbs the hydrostatic equilibrium \citep[e.g.][]{maeder2009}, or the properties of convective motions, depending on the ratio of convective to rotational velocities \citep[e.g.][]{kapyla2024}. Despite its importance, no conclusive description of the interplay between rotation and stellar evolution exists. Observations of internal rotation rates \citep{mosser2012, gehan2018, aerts2019, li2024} show that there must be a very efficient angular momentum transport mechanism from the core to the envelope as the stars evolve up the red giant branch (RGB) \citep[e.g.][]{eggenberger2012, eggenberger2017, ceillier2013, marques2013, spada2016}. To date, several mechanisms have been suggested to explain this angular momentum transport --- for example, hydrodynamical instabilities \citep{maeder2009}, internal gravity waves \citep{alvan2013, fuller2014, pincon2017}, mixed modes \citep{belkacem2015a, belkacem2015b, bordadagua2025} and magnetic fields \citep{spruit2002, cantiello2014, fuller2019, eggenberger2019,takahashi2021} --- but none are able to fully explain rotational evolution along the RGB. Measuring accurate internal rotation rates of red giants is therefore crucial for testing and constraining stellar evolution models.
	
	As in the case of the Sun, global oscillations in red giants are stochastically driven by turbulent convection in the outer layers of the star. This excitation mechanism drives both radial and non-radial oscillations over multiple overtones, allowing us to probe the internal structure of these stars. In red giants, we have the further advantage of being able to observe so-called mixed modes --- that is, modes that have a gravity ({\it g}-mode) character in the core and a pressure ({\it p}-mode) character in the outer layers of the star. The observation of these mixed modes allows us to probe the core properties, including core rotation rates, of red-giant stars.
	
	In the presence of rotation, the oscillation frequencies of non-radial mixed modes are perturbed. This perturbation can be measured from the splittings of these modes into multiplets in the oscillation power spectrum of red-giant stars, and thus facilitates estimation of internal rotation rates. In the linear regime, these frequency perturbations are the sum of rotational perturbations in the entire stellar interior. This makes them a cumulative measure of the internal rotation rates, depending both on the underlying rotation profile (which is unknown a priori) and the eigenfunction of the oscillation mode. For example, to determine core and envelope rotation rates, we need to disentangle the contribution of rotation from the different regions in the interior. Most commonly used methods combine information from a number of oscillation modes to construct localised estimates of the internal rotation profile at targeted regions within the star. Several different approaches have been pursued. Among these are the regularised least-squares \citep[RLS, see e.g. ][and references therein]{christensen1990}, optimally localised averages \citep[OLA, ][]{backus1968}, utilization of a linear relation between splittings and the mode trapping parameter $\zeta$ \citep{goupil2013, deheuvels2015, ahlborn2022}, and Markov chain Monte Carlo \citep[][]{fellay2021, buldgen2024}.
	
	All of these methods employ a linear relation between the rotational splittings on one hand, and the sensitivity function of the oscillation mode and the rotation profile on the other. However, this relation becomes non-linear in the presence of near-degeneracy effects \citep[NDE,][]{lynden1967, dziembowski1992, suarez2006, ouazzani2012, deheuvels2017,ong2022, li2024}. Such NDE occur when the rotation frequency of the star becomes comparable to the frequency differences between subsequent mixed modes, or the strength of the coupling between p- and g-modes. Hence, to use the above-mentioned methods, and linear rotational inversions in particular, we need to understand the regimes in which the assumption of linearity applies.

	In previous works that study the internal rotation rates of red-giant stars by means of rotational inversions, NDE have been neglected \citep{deheuvels2012, dimauro2016, dimauro2018, triana2017}, and no assessment of the potential systematic errors that are introduced by NDE is made. This necessitates a thorough systematic analysis of the applicability of linear rotational inversions as a function of the evolution on the RGB and the underlying rotation rates for these types of stars, which we carry out in this work.
	We computed synthetic rotational splittings for red-giant models taking the effects of NDE into account (see Sects.~\ref{secasy} and~\ref{secmethods}). We subsequently invert for these synthetic rotational splittings using linear inversion methods (see Sect.~\ref{secinv}). Given the known underlying rotation rates, we quantify the systematic errors in the estimated rotation rates. We show that linear rotational inversions are still applicable for less evolved stars or slower rotation rates, like it is the case for the red giant analysed in \cite{dimauro2016}. For faster rotating or more evolved red giants like in the sample of \cite{triana2017}, the assumption of linearity is no longer valid and errors introduced by NDE surpass the observational uncertainties (see Sect.~\ref{secresults}). We hence conclude that NDE effects need to be taken into account when interpreting estimated rotation rates of more evolved and faster rotating red-giant stars. We also identify ranges of stellar parameters that can be exploited by means of linear rotational inversions in future work.

	%%%%%%%%%%%%%%%%%%%%%%%%%%%%%%%%%%%%%%%%%%%%%
	\section{Rotational inversions}\label{secinv}
	The observation of stellar oscillation frequencies enables a probe of rotation inside of stars because the internal rotation of a star perturbs the non-radial mode frequencies (i.e. of modes with a spherical degree $\ell>0$). Each oscillation mode of a star can be characterised by three `quantum numbers': the radial order $n$, the spherical degree $\ell$, and the azimuthal order $m$. Without rotation, all modes of a given spherical degree $\ell$ and radial order $n$, but different $m$, have the same frequency (i.e. they are degenerate). In a rotating star, however, the modes with different $m$ have their degeneracy lifted such that we in principle can observe up to $2\ell+1$ frequency values, depending on the angle of inclination. The frequency difference between modes of subsequent $m$ is called the rotational splitting and is denoted as $\delta\omega_{n \ell m}$. 
	
	Most commonly-used rotational inversion methods neglect NDE and assume that the observed rotational splitting can be expressed as:
	\begin{align}
		\delta\omega_{n \ell m} = m\int_0^R\mathcal{K}_{n \ell}(r)\Omega(r)\text{d} r\equiv \delta\omega_{n\ell m,\,{\rm symm}},
		\label{eqsplitting}
	\end{align}
	where $\mathcal{K}_{n \ell}(r)$ denotes the so-called rotational kernel, which describes the sensitivity of the oscillation frequency to the rotation rate $\Omega(r)$ at radius $r$. For a more detailed discussion of the components of this equation we refer to Appendix~\ref{secperttheo}. Equation~\eqref{eqsplitting} relates the (symmetric) rotational splitting $\delta\omega_{n\ell m,\,{\rm symm}}$ linearly to the rotational kernel $\mathcal{K}_{n \ell}(r)$ and the internal rotation profile $\Omega(r)$. The rotational splitting is also linear in the azimuthal order, hence resulting in symmetric rotational splittings around the central ($m=0$) frequency of a multiplet. In Sect.~\ref{secasy} we discuss how the description of the rotational splitting changes when NDE are taken into account.
	
	In this work, we focus on OLA inversions. These allow us to construct a localised sensitivity function: the averaging kernel. To localise  this kernel at a target radius $r_0$, one uses a linear combination of the individual rotational kernels and a set of inversion coefficients $c_i(r_0)$:
	\begin{align}
		K(r, r_0) = \sum_{i\in\mathcal{M}}c_i(r_0)\mathcal{K}_{i}(r)\,,
		\label{eqavgkernel}
	\end{align}
	where $\mathcal{M}$ denotes the mode set of interest. In Eq.~\eqref{eqavgkernel} $i$ denotes the set of quantum numbers $(n,\ell)$.
	
	The inversion coefficients are chosen such that the averaging kernel is localised at the target radius. The quality of the localisation depends, however, on the underlying mode set \citep[e.g.][]{ahlborn2020}. An estimate of the rotation profile at the target radius ($\overline{\Omega}(r_0)$) can be computed as the average of the underlying rotation profile using the averaging kernel as the weighting function:
	\begin{align}
		\overline{\Omega}(r_0) = \int_0^RK(r, r_0)\Omega(r)\text{d}r\,.\label{eqomegaest}
	\end{align}
	If the averaging kernel were localised exactly at the target radius --- that is, if it was a Dirac $\delta$ distribution $K(r, r_0)=\delta(r-r_0)$ --- the average would be identical to the rotation rate at the target radius: $\overline{\Omega}(r_0)=\Omega(r_0)$. In practice, however, the sensitivity of $K(r,r_0)$ is spread over a range of radii around $r_0$, such that $\overline{\Omega}(r_0)$ is to be interpreted as the average value of $\Omega(r)$ over this range. Due to the linearity of Eq.~\eqref{eqsplitting}, Eq.~\eqref{eqomegaest} can be rewritten as:
	\begin{align}
		\overline{\Omega}(r_0) = \sum_{i\in\mathcal{M}}c_i(r_0)\delta\omega_{i}\,,
	\end{align}
	where $\delta\omega_{i}$ refers to the rotational splitting excluding the $m$ dependence: $\delta\omega_{i}=\delta\omega_{n\ell m}/m, (m\neq0)$.
	
	The determination of the inversion coefficients depends on the chosen inversion method. Depending on the objective function that is minimised to determine the inversion coefficients, two main types of OLA inversions have been used: subtractive \citep[SOLA, ][]{pijpers1992, pijpers1994} and multiplicative \citep[MOLA][]{backus1968} OLA inversions. Here, we use extended multiplicative OLA \citep[eMOLA, ][]{ahlborn2022, ahlborn2025} inversions as they have been shown to provide less biased estimates of red-giant envelope rotation rates, and yield values that are equivalent to those obtained from other OLA methods for core rotation rates. This is achieved by modifying the objective function of eMOLA such that it suppresses cumulative sensitivity away from the target radius.
	
	%%%%%%%%%%%%%%%%%%%%%%%%%%%%%%%%%%%%%%%%%%%%%
	\section{Near-degeneracy effects}\label{secasy}
	To describe the NDE we follow the discussions by \cite{deheuvels2017} and \cite{li2024} \citep[see also][]{dziembowski1992,suarez2006, ouazzani2012}. We make the Ansatz that the eigenfunctions $\boldsymbol{\xi}$ can be written as a linear combination of their unperturbed equivalents of the same $\ell$ and $m$ \citep{lynden1967, li2024}:
	\begin{align}
		\boldsymbol{\xi}(r,\theta,\phi,t)=\sum_i a_i\,\boldsymbol{\xi}_{0,i}(r,\theta,\phi,t)\,,\label{eqeigenfunction}
	\end{align}
	where $\boldsymbol{\xi}_{0,i}$ are the unperturbed eigenfunctions and $a_i$ are a set of contribution coefficients.
	
	To determine mode frequencies in the presence of NDE we need to solve the oscillation equation including the lowest-order perturbative effects of rotation \citep[e.g.][]{dyson1979,ong2022}:
	\begin{align}
		(\mathcal{L}+\omega\mathcal{R} - \omega^2\mathcal{D})\boldsymbol{\xi}=0\,.\label{eqosc}
	\end{align}
	Equation~\eqref{eqosc} is equivalent to Eq.~(1.1) in \citet{dyson1979} if one assumes that $\boldsymbol{\xi} \propto\exp(-\di\omega t)$, where $\omega$ is the mode frequency in the inertial frame, and retains terms that are first order in $\Omega$.
	This equivalence is shown in Appendix~\ref{secperteqmo}.
	The unperturbed operator $\mathcal{L}$ and the operator that describes the first order perturbation due to rotation $\mathcal{R}$ are given as
	\begin{align}
		\mathcal{L}\,\boldsymbol{\xi} &= \frac{\nabla p'}{\rho_0} + \nabla \Phi' - \dfrac{\rho'}{\rho_0^2} \nabla p_0\,,\label{eqop1}\\
		\mathcal{R}\,\boldsymbol{\xi}&=2(m\Omega-\di\boldsymbol{\Omega}\times)\boldsymbol{\xi}\,,\label{eqop2}
	\end{align}
	where $p, \rho$ and $\Phi$ refer to the pressure, density and gravitational potential, respectively. Equilibrium quantities are denoted with a subscript 0 while the Eulerian perturbations are denoted with a prime. The operator $\mathcal{D}$ is the standard inner product and the following rotation profile is assumed: $\mathbf{\Omega}=\Omega(r)(\cos\theta\,\mathbf{e}_r-\sin\theta\,\mathbf{e}_\theta)$ with $\mathbf{e}_r$ and $\mathbf{e}_\theta$ being the unit vectors of spherical coordinates.
	
	We now rewrite Eq.~\eqref{eqosc} as the following matrix equation:
	\begin{align}
		(\mathbf L+\omega\mathbf {R}-\omega^2\mathbf{D})\mathbf{a}=0\,,\label{eqoscmat}
	\end{align}
	where the perturbed mode eigenfrequencies are approximated as the eigenvalues of this equation. The corresponding eigenvectors $\mathbf{a}$ describe the contributions of the unperturbed eigenfunctions to the eigenfunction under the presence of NDE (see Eq.~\eqref{eqeigenfunction}). The matrix elements of the different matrices in Eq.~\eqref{eqoscmat} are defined as:
	\begin{subequations}
		\begin{align}
			L_{ij}&=\langle\boldsymbol{\xi}_{0,i}\,|\,\mathcal{L}\boldsymbol{\xi}_{0,j}\rangle=\delta_{ij}\cdot\omega_{0,i}^2\,,\\
			R_{ij}&=\langle\boldsymbol{\xi}_{0,i}\,|\,\mathcal{R}\boldsymbol{\xi}_{0,j}\rangle=2\delta\omega_{ij}\,,\\ \label{eqdij}
			D_{ij}&=\langle\boldsymbol{\xi}_{0,i}\,|\,\boldsymbol{\xi}_{0,j}\rangle = \delta_{ij}\,,
		\end{align}
	\end{subequations}
	where $\omega_{0, i}$ are the unperturbed eigenfrequencies and $\delta\omega_{ij}$ are the frequency perturbations introduced to mode $i$ (with ($n,\ell,m$)) by the rotational interaction with mode $j$ ($n',\ell,m$). The observable rotational splitting of mode $i$ is then a combination of all the frequency perturbations $\delta\omega_{ij}$ which can only be obtained by solving Eq.~\eqref{eqoscmat}. The inner product of two eigenfunctions $\boldsymbol{\xi}_a,\boldsymbol{\xi}_b$ is written with angle brackets and is defined as:
	\begin{align}
		\langle \boldsymbol{\xi}_a\,|\,\boldsymbol{\xi}_b\rangle =\int_V\rho_0\,\boldsymbol{\xi}_a^*\cdot\boldsymbol{\xi}_b\,{\rm d}V\,. \label{eqinnerproduct}
	\end{align}
	We also assume orthonormal eigenfunctions  such that $\langle\boldsymbol{\xi}_{0,i}\,|\,\boldsymbol{\xi}_{0,j}\rangle=\delta_{ij}$, where $\delta_{ij}$ denotes the Kronecker $\delta$. The frequency perturbations $\delta\omega_{ij}$ introduced by rotation are computed as
	\begin{align}
		\delta\omega_{ij} = m \int_0^R\mathcal{K}_{ij}(r)\,\Omega(r)\text{d} r\,,
		\label{eqcrosssplitting}
	\end{align}
	where $\mathcal{K}_{ij}$ is given by Eq.~\eqref{eqcrosskernel}. We note that for $i=j$ this is identical to the commonly used rotational kernels, defined in Eq.~\eqref{eqsplitting}. The rotational perturbations of the mode frequencies also include off-diagonal coupling terms in $R_{ij}$, because of the NDE.
	In the far-resonance limit, these off-diagonal terms account for second-order rotational frequency corrections \citep[see e.g.][who we shall hereafter refer to as \citetalias{dziembowski1992}, and who discuss NDE for 2 modes]{dziembowski1992}.
	We describe in Appendix~\ref{secperttheo} how to practically compute the lowest-order frequency perturbations introduced by rotation.
	
	Equation~\eqref{eqoscmat} constitutes a quadratic eigenvalue problem (QEP). To solve this QEP, we linearise Eq.~\eqref{eqoscmat} to yield the generalised eigenvalue problem $\mathbf A\mathbf{a}'=\omega \mathbf B\mathbf{a}'$ with newly defined matrices $\mathbf{A},\mathbf{B}$ and eigenvectors $\mathbf{a}'$. For more details on how we solve this QEP we refer to Appendix~\ref{secqep}. In addition, we applied other solution schemes for solving Eq.~\eqref{eqoscmat}, for example using the \cite{deheuvels2017} linearised equations as well as what we call the generalised \citetalias{dziembowski1992} formalism described in Appendix~\ref{sectilde}. We find that the solution of the generalised eigenvalue problem and the alternative approaches yield almost identical results in terms of the rotational splitting asymmetries (see e.g. Fig.~\ref{figasymmetrycomparison} for results obtained with the generalised \citetalias{dziembowski1992} formalism).
	
	The eigenvectors of this eigenvalue problem provide us with the contribution coefficients $a_i$ to compute the near-degenerate eigenfunctions in Eq.~\eqref{eqeigenfunction}. We note that NDE are only expected to occur between modes that are close in frequency, and are hence strongest among neighbouring frequencies. This is reflected in decreasing contribution coefficients $a_i$ for increasing frequency differences and, in perturbation theory, this is reflected by the fact that expansions for these $a_i$ are in powers of not just $\delta\omega_{ij}$, but rather $\delta\omega_{ij} / (\omega^2_i - \omega^2_j)$. We confirm this by numerically solving the eigenvalue problem described by Eq.~\eqref{eqoscmat}. The total contribution due to the interaction with other modes (i.e. contribution of the off-diagonal elements) is largest for {\it p}-dominated modes. This is expected due to the smaller frequency differences between subsequent mixed modes.
	
	%%%%%%%%%%%%%%%%%%%%%%%%%%%%%%%%%%%%%%%%%%%%%
	\section{Synthetic data}\label{secmethods}
	In order to explore the applicability of the linear inversion method discussed in Sect.~\ref{secinv}, we confront it with different sets of synthetic data. Using the methods described in Sect.~\ref{secasy}, we computed rotational splittings in the presence of NDE.
	\begin{figure*}
		\centering
		\begin{subfigure}{.5\textwidth}
			\centering
			\includegraphics[]{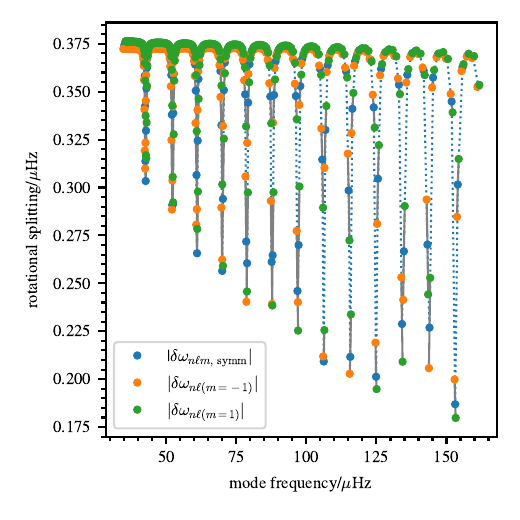}
		\end{subfigure}%
		\begin{subfigure}{.5\textwidth}
			\centering
			\includegraphics[]{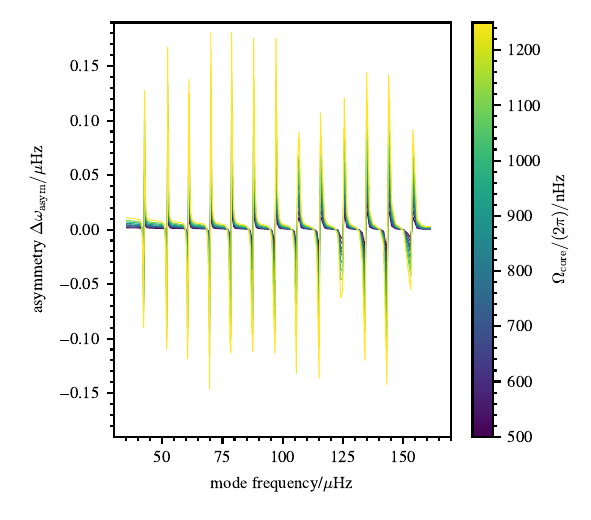}
		\end{subfigure}
		\caption{Rotational splittings and their asymmetries as a function of mode frequency. {\it Left panel:} Rotational splittings of the evolved model using a core rotation rate of $\Omega_{\rm core}=750$~nHz and an envelope rotation rate of $\Omega_{\rm env}=50$~nHz. Rotational splittings that have the same $n_{pg}$ are connected with a grey line. We note that we select only a subset of the rotational splittings centred around $\nu_{\rm max}\approx100~\mu$Hz for the rotational inversions (see text for details). The dotted, blue line connects the symmetric splittings for the sake of readability. The rotational splittings $\delta\omega_{n\ell m}$ were obtained by computing the frequency differences according to Eq.~\eqref{eqsplittingsasym1} and \eqref{eqsplittingsasym2} using the frequencies obtained from solving the full QEP posed by Eq.~\eqref{eqoscmat} as described in Appendix~\ref{secqep}, while the symmetric rotational splittings $\delta\omega_{n\ell m,\, {\rm symm}}$ where obtained from Eq.~\eqref{eqsplitting}. {\it Right panel:} Asymmetries of the rotational splittings ($\Delta\omega_{\rm asym}$) of the evolved model over a range of core rotation rates $\Omega_{\rm core}$ and for $\Omega_{\rm env}=50$~nHz. The lines are colour-coded by the core rotation rate.}
		\label{figasymsplittings}
	\end{figure*}
	
	For the computation of the synthetic data, we used an evolutionary track of a 1~M$_\odot$ star with solar metallicity \citep{asplund2009} constructed with the Modules for Experiments in Stellar Astrophysics \citep[MESA, r12778,][ and references therein]{paxton2011,paxton2013,paxton2015,paxton2018, paxton2019, jermyn2023} stellar evolution code. We used the {\tt mesa} equation of state and OPAL opacities \citep{iglesias1996} extended with low temperature opacities by \cite{ferguson2005}. Convection was treated using the mixing-length theory \citep{boehm1958} with a mixing-length parameter of $\alpha_{\rm MLT}=1.8$. In the following sections we discuss a model with a large frequency separation of $\Delta\nu\approx9~\mu$Hz and a frequency of maximum oscillation power $\nu_{\rm max}\approx100~\mu$Hz from this evolutionary track in more detail. We refer to this model as the `evolved model'. We note that this model is identical to the `$\Delta\nu=9~\mu$Hz model' in \cite{ahlborn2025}. To compute the large frequency separation $\Delta\nu$ we used scaling relations \citep{kjeldsen1995} with reference values taken from \cite{themessl2018} and the mass and radius of the stellar models. We computed oscillation frequencies and eigenfunctions using the stellar oscillation code GYRE \citep[version 7.1,][]{townsend2013, townsend2018}. For the generation of the synthetic mode sets, including synthetic uncertainties, we follow \cite{ahlborn2025}. Each mode set contains 12 dipole modes across four radial orders centred around $\nu_{\rm max}$. For each radial order, we selected the most {\it p}-dominated mode and two of the least {\it g}-dominated modes. 
	
	To compute the synthetic rotational splittings and the NDE we need to assume a synthetic rotation profile. We used a profile that features a step at the base of the convective envelope and has otherwise constant rotation rates (i.e. a two-zone model such as those defined in \citealt{klion2017,ahlborn2020}). In the following sections, we assume varying ratios of the core to envelope rotation rates. We computed the rotational splittings for $m=\pm 1$ as the differences of the frequencies obtained when solving the full QEP posed by Eq.~\eqref{eqoscmat} with different $m$  as described in Appendix~\ref{secqep}:
	\begin{subequations}
		\begin{align}
			\delta\omega_{n\ell( m=-1)} &= \omega_{n\ell( m=-1)}-\omega_{n\ell( m=0)}\,,\label{eqsplittingsasym1}\\
			\delta\omega_{n\ell( m=1)} &= \omega_{n\ell( m=1)}-\omega_{n\ell( m=0)}\,.\label{eqsplittingsasym2}
		\end{align}
	\end{subequations}
	Symmetric rotational splittings $\delta\omega_{n\ell m,\, {\rm symm}}$ for comparison were computed using Eq.~\eqref{eqsplitting}. We illustrate the rotational splittings computed for $m=\pm 1$ in the left panel of Fig.~\ref{figasymsplittings}. For the chosen rotation rates, the differences between the splittings of different azimuthal order can be clearly seen. To illustrate the difference between the splittings of different $m$ we define the asymmetry of the splittings as
	\begin{align}
		\Delta\omega_{\rm asym} &= \delta\omega_{n\ell( m=1)}+\delta\omega_{n\ell( m=-1)}\label{eqasymmetry}\\
		&\equiv\omega_{n\ell( m=-1)}+\omega_{n\ell( m=1)}-2\omega_{n\ell( m=0)}\,.\nonumber
	\end{align}
	This asymmetry is shown in the right panel of Fig.~\ref{figasymsplittings} for varying core rotation rates. The panel clearly shows how the asymmetry increases for increasing core rotation rates. We note that in the case of a core magnetic field the frequency perturbations would have the same sign for modes with different $m$. Therefore, the resulting asymmetries on the rotational splittings, would not change sign as a function of frequency \citep[see][their Fig. 2]{li2022}. Further, magnetic frequency perturbations are expected to be higher for the {\it g}-dominated modes, as the magnetic field is typically assumed to be located in the core \citep[e.g.][]{loi2021, bugnet2021}. These features can be observationally distinguished from the pattern of asymmetries introduced by rotationally induced NDE shown in the right panel of Fig.~\ref{figasymsplittings}.
	
	As an input for the rotational inversions we compute `averaged' dipole rotational splittings:
	\begin{align}
		\delta\omega_{n\ell,\,{\rm avg}} &= \frac{1}{2}(\delta\omega_{n\ell( m=1)}-\delta\omega_{n\ell( m=-1)})\label{eqsplittingsymm}\\
		&\equiv\frac{1}{2}(\omega_{n\ell( m=1)}-\omega_{n\ell( m=-1)})\,,\nonumber
	\end{align}
	because the linear inversions discussed in Sect.~\ref{secinv} assume a symmetric splitting following Eq.~\eqref{eqsplitting}. A similar approach could be adopted for observed rotational splittings, especially in cases where no $m=0$ component is visible. We invert for this set of averaged rotational splittings using the rotational inversions and compare the estimated rotation rates to the known input rotation rates (see Sect.~\ref{secresults}). We note that the difference between the averaged splittings and the splittings for different $m$ is half the asymmetry $\Delta\omega_{\rm asym}$.
	
	As an independent confirmation of the NDE formalism, we also computed oscillation frequencies using a modified version of the $\pi$-$\gamma$ formalism \citep{ong2020,ong2021}. The details are discussed in Appendix~\ref{secpigamma}. We find good agreement in terms of shape and magnitude of the asymmetries between both formalisms (see Fig.~\ref{figasymmetrycomparison}). We consider this as an important confirmation of the robustness of our synthetic data.
	%%%%%%%%%%%%%%%%%%%%%%%%%%%%%%%%%%%%%%%%%%%%%
	\section{Impact of near degeneracy effects}\label{secresults}
	The NDE come into play when the frequency difference between the unperturbed modes is of the order of the rotational splittings, which is proportional to the rotation rate of the star; that is, $|\omega_{0, i}-\omega_{0, j}|\sim\Omega$ (see \citetalias{dziembowski1992}; \citealt{deheuvels2017}; \citealt{li2024}). It is hence expected that the error introduced by the NDE in the linear inversion result is a function of the underlying rotation rate of the star. In addition, the impact of NDE is also expected to increase along evolution on the RGB \citep{deheuvels2017, li2024}. As the stars evolve along the RGB, their cores continue to contract while their envelopes expand. As an overall effect, the mixed mode density increases (and the coupling strength decreases) with evolution up the red giant branch, leading to smaller frequency differences between the mixed modes in each acoustic radial order, and between p- and g-dominated mixed modes. Hence, the impact of NDE on the rotational inversion results becomes more important in later evolutionary stages. In this section we first explore the dependence on the position of the star along the RGB and subsequently discuss how the impact of NDE changes with varying rotation rates for a more evolved model with $\Delta\nu\approx9~\mu$Hz. For the remainder of this section, we estimate core ($\Omega_{\rm core}$) and envelope  ($\Omega_{\rm env}$) rotation rates by constructing averaging kernels at  $r_0=0.003$ and 0.98, respectively. To disentangle errors introduced by the NDE from any other error (e.g. structural differences changing the mixing fractions and thus the rotational kernels --- see \citealt{ong2024}) we invert for the synthetic rotational splittings by using the stellar model for which we generated the synthetic data as a reference model.
	%%%%%%%%%%%%%%%%%%%%%%%%%%%%%%%%%%%%%%%%%%%%%%%%%%%%%%%%%%%%%%
	\subsection{Lower RGB}
\begin{figure}
		\centering
		\includegraphics[]{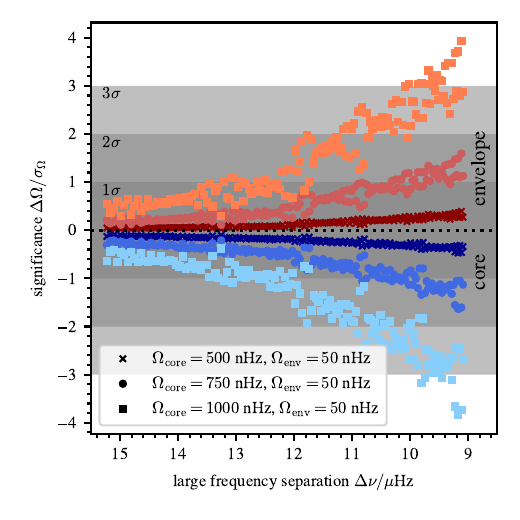}
		\caption{Error on the estimated core and envelope rotation rates introduced by neglecting the NDE. The errors are shown in terms of the individual uncertainties $\sigma_\Omega$ as a function of the large frequency separation $\Delta\nu$. We generated synthetic splittings for three different core rotation rates of $\Omega_{\rm core}=500, 750, 1000$~nHz and a envelope rotation rate of $\Omega_{\rm env}=50$~nHz indicated with crosses, circles and squares, respectively. Results for the core and envelope are shown in different shades of blue and red, respectively. Note that the models evolve from left to right in this figure.}
		\label{figsignificance}
	\end{figure}

	To explore how the impact of NDE varies along the lower RGB, we computed asymmetric rotational splittings along an evolutionary track using our forward model described in Sects.~\ref{secasy} and~\ref{secmethods}. Subsequently, we inverted for the core and envelope rotation rate using the averaged rotational splittings computed according to Eq.~\eqref{eqsplittingsymm}. The results in Fig.~\ref{figsignificance} clearly show that for the least evolved stars NDE can be neglected, whereas for the most evolved stars considered, NDE may lead to significantly overestimated envelope rotation rates and significantly underestimated core rotation rates.
	
	In Fig.~\ref{figsignificance} the models at the high end of $\Delta\nu$ are similar to the star studied in \cite{dimauro2016, dimauro2018} while models at the low end are more similar to the stars in the sample of \cite{triana2017}. The range of models in Fig.~\ref{figsignificance} hence covers the parameters of red-giants with estimated envelope rotation rates that are currently available. We have chosen three representative values of the core rotation rate (500, 750 and 1000~nHz) and one representative value of the envelope rotation rate of 50~nHz.\footnote{These values cover the ranges of observed core and envelope rotation rates. In Sect.~\ref{secrotrate} we show that the errors decrease with increasing envelope rotation rate. Hence, we do not show results for larger envelope rotation rates in this section.}
	
	The deviation for the estimated envelope rotation rate is positive while it is negative for the estimated core rotation rate along the evolution on the RGB --- that is, the inversion procedure systematically overestimates envelope rotation rates, and systematically underestimates core rotation rates. We note that observational uncertainties on the estimated envelope rotation rate range from approximately 12~nHz at the high end to 21~nHz at the low end of our $\Delta\nu$ range, whereas the uncertainties of estimated core rotation rates vary from 12~nHz to 16~nHz over the same range. Hence, a deviation of 2$\sigma$ or larger becomes very quickly comparable to the underlying value of the envelope rotation rate and makes the result difficult to interpret. This is less problematic for the estimated core rotation rates as the underlying values are typically ten times higher than those of the envelope rotation rates. The errors due to NDE also become smaller with decreasing core rotation rate, as is expected.
	
	Based on these results, we identified ranges of core rotation rates and large frequency separations in which we consider the linear inversions to still apply for the given test case. For the case with a core rotation rate of 500~nHz or lower, NDE do not seem to play a significant role along evolution in the ranges considered here. For the intermediate case of 750~nHz the estimated rotation rates become discrepant beyond $1\sigma$ below a value of $\Delta\nu\approx11~\mu$Hz. For the high core rotation rate of 1000~nHz this happens earlier, at a value of $\Delta\nu\approx13~\mu$Hz. Considering previous estimates of core rotation rates \citep{mosser2012,triana2017,gehan2018,li2024} the transition from reliable estimates of the envelope rotation rate to unreliable estimates resides right within the observed ranges of $\Delta\nu$ and $\Omega_{\rm core}$. This highlights the importance of NDE when interpreting linear rotational inversion results on the RGB.
	%%%%%%%%%%%%%%%%%%%%%%%%%%%%%%%%%%%%%%%%%%%%%%%%%%%%%%%%%%%%%%
	\subsection{Evolved model with $\Delta\nu=9~\mu$Hz}\label{secrotrate}
	\begin{figure*}
		\centering
		\begin{subfigure}{.5\textwidth}
			\centering
			\includegraphics[]{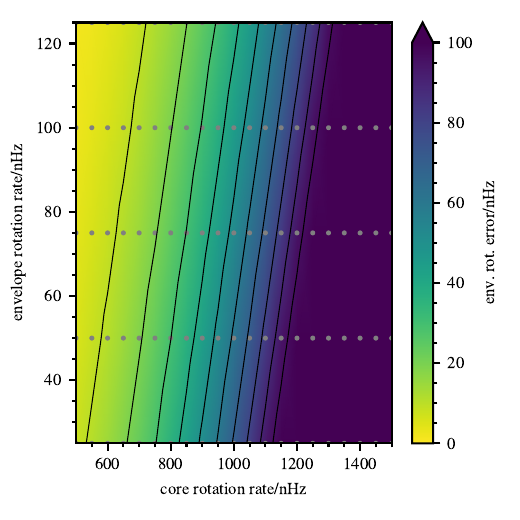}
		\end{subfigure}%
		\begin{subfigure}{.5\textwidth}
			\centering
			\includegraphics[]{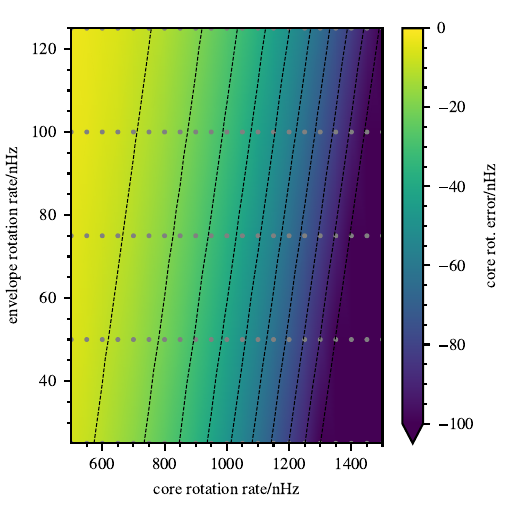}
		\end{subfigure}
		\caption{Absolute error of the estimated envelope and core rotation rates for the evolved model. {\it Left panel:} The absolute error of the estimated envelope rotation rate as a function of input core and envelope rotation rate is shown. The input core and envelope rotation rates are given on the $x$- and $y$-axis, respectively. The error of the estimated envelope rotation rate in terms of nHz is colour coded and cut at a maximum value of 100~nHz. The grey dots indicate the underlying grid of input core and envelope rotation rates. The contour lines have a separation of 10~nHz. {\it Right panel:} Same as the left panel now for the absolute error of the estimated core rotation rate. Note that the colour scale is inverted, as the errors on the core rotation rate are negative.}
		\label{figevolvedsurf}
	\end{figure*}
	In this section, we explore the impact of NDE on the inversion results for a single stellar model at the low $\Delta\nu$ end in Fig.~\ref{figsignificance} in more detail \citep[identical to the `$\Delta\nu=9~\mu$Hz model' in Table 1 of][see also Sect.~\ref{secmethods}]{ahlborn2025}. We show the absolute error of the estimated envelope and core rotation rates for varying input core and envelope rotation rates in Fig.~\ref{figevolvedsurf}, respectively. These results show that it is possible to recover core and envelope rotation rate with small errors within certain limits of the underlying rotation rates even for more evolved stars. This is equivalent to assuming that the rotational splittings can be approximated by Eq.~\eqref{eqsplitting} for low enough rotation rates.
	
	The left panel of Fig.~\ref{figevolvedsurf} reveals two immediate trends: (i) at fixed envelope rotation rate the absolute error of the estimated envelope rotation rate increases with increasing core rotation rate, (ii) for a fixed core rotation rate the absolute error decreases with increasing envelope rotation rate. That is, the higher the ratio of the core-to-envelope rotation rates, the higher the error introduced on the estimated envelope rotation rate. However, the dependence of this absolute error on the input envelope rotation rate is much less strong than its dependence on the input core rotation rate: for an increase of the envelope rotation rate by a factor of five, the error approximately decreases by a factor of two to four depending on the core rotation, while the error increases approximately by a factor of ten when doubling the core rotation rate. This can be explained by the fact that the core rotates ten times faster than the envelope. A doubling of the core rotation rate thus corresponds to a much larger absolute difference, indicating that the error of the estimated envelope rotation rate is dominated by the underlying core rotation rate.
	
	As discussed above, the impact of the NDE depends on the frequency differences between subsequent modes (i.e. modes with the same $\ell$ and $m$ but different $n$) in relation to the rotational splitting of the modes. In addition to the values of the core and envelope rotation rates, the rotational splitting also depends on the shape of the underlying rotation profile. Here, we have to consider that in most red giants the {\it g}-modes are well described in the Jeffreys-Wentzel-Kramers-Brillouin (JWKB) regime, such that basically all their {\it g}-modes sense the same rotation rate. In the case of a rotation profile with a step at the base of the convection zone (used here), the whole {\it g}-mode cavity is rotating at a constant rate. In a case with a more centrally concentrated rotation profile, for example with a transition around the hydrogen burning shell, the rotation rate changes within the {\it g}-mode cavity. This leads to smaller rotational splittings as compared to the previous case because the {\it g}-mode component becomes sensitive to the more slowly rotating envelope \citep[e.g.][]{ahlborn2022, ahlborn2025}. However, from the perspective of NDE, this is indistinguishable from the case where the core were set to rotate uniformly at an effective rotation rate (averaged over the {\it g}-mode cavity) instead. Thus, for the same reason that NDE are less significant with slower-rotating cores, NDE are also less significant with more centrally concentrated rotational profiles.
	
	The absolute errors for the estimated core rotation rates are shown in the right panel of Fig.~\ref{figevolvedsurf}. The trends are qualitatively the same as for the estimated envelope rotation rate. Contrary to our findings for the envelope rotation rates, core rotation rates are systematically underestimated when neglecting NDE. Overall, the absolute errors of the core rotation rate are smaller than those of the envelope rotation rate. They do, however, remain of comparable magnitude. Considering that the core rotates about ten times faster than the envelope, the relative errors of the estimated rotation rate of the core are much smaller than those estimated for the envelope. From an analysis of Fig.~\ref{figevolvedsurf} we conclude, that for the model under consideration, there is a range of core rotation rates, within which we can neglect the effects of the NDE and use linear rotational inversions. For high core rotation rates, however, this reinforces our earlier conclusion that envelope rotation rates are systematically overestimated, while core rotation rates are systematically underestimated.
	
	%--------------------------------------------------------------------
	\section{Conclusions}\label{seconclusions}
	We showed that NDE may have a significant impact on the rotational inversion results of red-giant stars. This necessitates a careful interpretation of the estimated core and envelope rotation rates estimated from techniques that inherently assume symmetric rotational splittings. However, we also show that within certain limits of the internal rotation rates, linear rotational inversions are still applicable along the RGB. In this work, we used the eMOLA method, which has been shown to typically deliver the most accurate envelope rotation rates. Considering the similarities with MOLA and SOLA inversions \citep[e.g.][]{ahlborn2022, ahlborn2025}, and especially the use of Eq.~\eqref{eqsplitting}, we conclude that our results present a lower limit for the estimated envelope rotation rates.
	
	Independent of the underlying rotation rates and position along the RGB, we find that the errors introduced in the estimated core and envelope rotation rates are of similar magnitude, and have opposite signs (positive for the envelope, negative for the core). As the core of red giants typically rotates ten to twenty times faster than the envelope, the relative error on the estimated envelope rotation rate is much larger than on the core rotation rate. Relatively speaking, this makes estimated core rotation rates more reliable than estimated envelope rotation rates: an error of 100~nHz on an estimated value of 1000~nHz still allows us to conclude that the core is rotating fast, while the same error on an estimate of 50~nHz renders reliable conclusions impossible. We can hence conclude that the estimated core rotation rates in the presence of NDE can be still used to judge whether NDE play a role.
	
	We further demonstrate that the impact of NDE does strongly vary with evolution. As shown in Fig.~\ref{figsignificance}, rotationally-induced NDE can be safely neglected for stars towards the base of the RGB, as was for example analysed in \cite{dimauro2016}. NDE from rotation alone are thus not the dominant source of errors for their linear rotational inversion results; cf. \citet{ong2024} for other sources of error. This remains true for the evolution along the lower RGB for the lowest core rotation rate considered. For the higher core rotation rates of 750 and 1000~nHz, NDE start to become significant ($>1\sigma$) at a $\Delta\nu$ of approximately $11~\mu$Hz and $13~\mu$Hz, respectively. This implies that the estimated envelope rotation rates of \cite{triana2017} are most likely biased towards higher absolute values for the stars with high estimated core rotation rates. Finally, we show that the error of the estimated rotation rates introduced by the NDE strongly increases with increasing core rotation rate. This confirms earlier results by \cite{deheuvels2017,li2024}.
	
	The conclusions on the impact of NDE are not exclusive to rotational inversions. Any method relying on Eq.~\eqref{eqsplitting} assumes a linear relation between the rotational splitting and the structural perturbation due to rotation. For example, RLS inversions using Bayesian approaches \citep[e.g.][]{beck2014, fellay2021} relate rotational splittings, computed using a forward model, to either a frequency list or the power spectrum, in their evaluation of likelihood functions. While the star analysed in \cite{fellay2021} is still low on the RGB --- where NDE are not expected to play a role --- the star studied in \cite{beck2014} has evolved further up the RGB, and in that case NDE could have introduced systematic errors. Such forward models hence need to take NDE into account to obtain accurate estimates of the core and envelope rotation rates.
	
	Disregarding the difficulties of estimating envelope rotation rates in the presence of high core rotation rates due to NDE, we find ranges in stellar parameters in which the envelope rotation rates can still be recovered within the $1\sigma$ uncertainties. These ranges encompass less evolved stars over a broader range of core rotation rates and the lowest input core rotation rates for the more evolved model considered in Fig.~\ref{figevolvedsurf}. We therefore expect that there are still a number of {\it Kepler} red-giants in the parameter ranges determined here that can be analysed using linear inversions methods (e.g. eMOLA). Exploiting this potential will be the task of future work.
	
	\begin{acknowledgements}
		We would like to thank the anonymous referee for helpful comments to improve this manuscript. The research leading to the presented results has received funding from the European Research Council under the European Community’s Horizon 2020 Framework/ERC grant agreement no 101000296 (DipolarSounds). We thank the Klaus Tschira foundation for their support. JMJO acknowledges support from NASA through the NASA Hubble Fellowship grant HST-HF2-51517.001, awarded by STScI. STScI is operated by the Association of Universities for Research in Astronomy, Incorporated, under NASA contract NAS5-26555. SB acknowledges NSF grant AST-2205026. She would like to thank the Heidelberg Institute of Theoretical Studies for their hospitality at the time that this project was conceived. Throughout this work we have made use of the following Python packages: {\sc SciPy} \citep{2020SciPy-NMeth}, {\sc NumPy} \citep{2020NumPy-Array}, {\sc Pandas} \citep{mckinney-proc-scipy-2010}, and {\sc Matplotlib} \citep{4160265}. We thank their authors for making these great software packages open source.
	\end{acknowledgements}
	
	\bibliographystyle{aa} % style aa.bst
	\bibliography{bibliography.bib} % your references Yourfile.bib

\begin{thebibliography}{75}
\expandafter\ifx\csname natexlab\endcsname\relax\def\natexlab#1{#1}\fi

\bibitem[{{Aerts} {et~al.}(2010){Aerts}, {Christensen-Dalsgaard}, \&
  {Kurtz}}]{aerts2010}
{Aerts}, C., {Christensen-Dalsgaard}, J., \& {Kurtz}, D.~W. 2010,
  {Asteroseismology} (Springer Dordrecht)

\bibitem[{{Aerts} {et~al.}(2019){Aerts}, {Mathis}, \& {Rogers}}]{aerts2019}
{Aerts}, C., {Mathis}, S., \& {Rogers}, T.~M. 2019, \araa, 57, 35

\bibitem[{{Ahlborn} {et~al.}(2020){Ahlborn}, {Bellinger}, {Hekker}, {Basu}, \&
  {Angelou}}]{ahlborn2020}
{Ahlborn}, F., {Bellinger}, E.~P., {Hekker}, S., {Basu}, S., \& {Angelou},
  G.~C. 2020, \aap, 639, A98

\bibitem[{{Ahlborn} {et~al.}(2022){Ahlborn}, {Bellinger}, {Hekker}, {Basu}, \&
  {Mokrytska}}]{ahlborn2022}
{Ahlborn}, F., {Bellinger}, E.~P., {Hekker}, S., {Basu}, S., \& {Mokrytska}, D.
  2022, \aap, 668, A98

\bibitem[{{Ahlborn} {et~al.}(2025){Ahlborn}, {Bellinger}, {Hekker}, {Basu}, \&
  {Mokrytska}}]{ahlborn2025}
{Ahlborn}, F., {Bellinger}, E.~P., {Hekker}, S., {Basu}, S., \& {Mokrytska}, D.
  2025, \aap, 693, A274

\bibitem[{{Alvan} {et~al.}(2013){Alvan}, {Mathis}, \& {Decressin}}]{alvan2013}
{Alvan}, L., {Mathis}, S., \& {Decressin}, T. 2013, \aap, 553, A86

\bibitem[{{Asplund} {et~al.}(2009){Asplund}, {Grevesse}, {Sauval}, \&
  {Scott}}]{asplund2009}
{Asplund}, M., {Grevesse}, N., {Sauval}, A.~J., \& {Scott}, P. 2009, \araa, 47,
  481

\bibitem[{{Backus} \& {Gilbert}(1968)}]{backus1968}
{Backus}, G. \& {Gilbert}, F. 1968, Geophysical Journal, 16, 169

\bibitem[{{Beck} {et~al.}(2014){Beck}, {Hambleton}, {Vos}, {Kallinger},
  {Bloemen}, {Tkachenko}, {Garc{\'{\i}}a}, {{\O}stensen}, {Aerts}, {Kurtz}, {De
  Ridder}, {Hekker}, {Pavlovski}, {Mathur}, {De Smedt}, {Derekas}, {Corsaro},
  {Mosser}, {Van Winckel}, {Huber}, {Degroote}, {Davies}, {Pr{\v s}a},
  {Debosscher}, {Elsworth}, {Nemeth}, {Siess}, {Schmid}, {P{\'a}pics}, {de
  Vries}, {van Marle}, {Marcos-Arenal}, \& {Lobel}}]{beck2014}
{Beck}, P.~G., {Hambleton}, K., {Vos}, J., {et~al.} 2014, \aap, 564, A36

\bibitem[{{Belkacem} {et~al.}(2015{\natexlab{a}}){Belkacem}, {Marques},
  {Goupil}, {Mosser}, {Sonoi}, {Ouazzani}, {Dupret}, {Mathis}, \&
  {Grosjean}}]{belkacem2015b}
{Belkacem}, K., {Marques}, J.~P., {Goupil}, M.~J., {et~al.} 2015{\natexlab{a}},
  \aap, 579, A31

\bibitem[{{Belkacem} {et~al.}(2015{\natexlab{b}}){Belkacem}, {Marques},
  {Goupil}, {Sonoi}, {Ouazzani}, {Dupret}, {Mathis}, {Mosser}, \&
  {Grosjean}}]{belkacem2015a}
{Belkacem}, K., {Marques}, J.~P., {Goupil}, M.~J., {et~al.} 2015{\natexlab{b}},
  \aap, 579, A30

\bibitem[{{B{\"o}hm-Vitense}(1958)}]{boehm1958}
{B{\"o}hm-Vitense}, E. 1958, \zap, 46, 108

\bibitem[{{Bordad{\'a}gua} {et~al.}(2025){Bordad{\'a}gua}, {Ahlborn},
  {Copp{\'e}e}, {Marques}, {Belkacem}, \& {Hekker}}]{bordadagua2025}
{Bordad{\'a}gua}, B., {Ahlborn}, F., {Copp{\'e}e}, Q., {et~al.} 2025, \aap,
  699, A310

\bibitem[{{Bugnet} {et~al.}(2021){Bugnet}, {Prat}, {Mathis}, {Astoul},
  {Augustson}, {Garc{\'\i}a}, {Mathur}, {Amard}, \& {Neiner}}]{bugnet2021}
{Bugnet}, L., {Prat}, V., {Mathis}, S., {et~al.} 2021, \aap, 650, A53

\bibitem[{{Buldgen} {et~al.}(2024){Buldgen}, {Fellay}, {B{\'e}trisey},
  {Deheuvels}, {Farnir}, \& {Farrell}}]{buldgen2024}
{Buldgen}, G., {Fellay}, L., {B{\'e}trisey}, J., {et~al.} 2024, \aap, 689, A307

\bibitem[{{Cantiello} {et~al.}(2014){Cantiello}, {Mankovich}, {Bildsten},
  {Christensen-Dalsgaard}, \& {Paxton}}]{cantiello2014}
{Cantiello}, M., {Mankovich}, C., {Bildsten}, L., {Christensen-Dalsgaard}, J.,
  \& {Paxton}, B. 2014, The Astrophysical Journal, 788, 93

\bibitem[{{Ceillier} {et~al.}(2013){Ceillier}, {Eggenberger}, {Garc{\'{\i}}a},
  \& {Mathis}}]{ceillier2013}
{Ceillier}, T., {Eggenberger}, P., {Garc{\'{\i}}a}, R.~A., \& {Mathis}, S.
  2013, \aap, 555, A54

\bibitem[{{Christensen-Dalsgaard} {et~al.}(1990){Christensen-Dalsgaard},
  {Schou}, \& {Thompson}}]{christensen1990}
{Christensen-Dalsgaard}, J., {Schou}, J., \& {Thompson}, M.~J. 1990, Monthly
  Notices of the Royal Astronomical Society, 242, 353

\bibitem[{{Deheuvels} {et~al.}(2015){Deheuvels}, {Ballot}, {Beck}, {Mosser},
  {{\O}stensen}, {Garc{\'{\i}}a}, \& {Goupil}}]{deheuvels2015}
{Deheuvels}, S., {Ballot}, J., {Beck}, P.~G., {et~al.} 2015, Astronomy and
  Astrophysics, 580, A96

\bibitem[{{Deheuvels} {et~al.}(2012){Deheuvels}, {Garc{\'{\i}}a}, {Chaplin},
  {Basu}, {Antia}, {Appourchaux}, {Benomar}, {Davies}, {Elsworth}, {Gizon},
  {Goupil}, {Reese}, {Regulo}, {Schou}, {Stahn}, {Casagrande},
  {Christensen-Dalsgaard}, {Fischer}, {Hekker}, {Kjeldsen}, {Mathur}, {Mosser},
  {Pinsonneault}, {Valenti}, {Christiansen}, {Kinemuchi}, \&
  {Mullally}}]{deheuvels2012}
{Deheuvels}, S., {Garc{\'{\i}}a}, R.~A., {Chaplin}, W.~J., {et~al.} 2012, The
  Astrophysical Journal, 756, 19

\bibitem[{{Deheuvels} {et~al.}(2017){Deheuvels}, {Ouazzani}, \&
  {Basu}}]{deheuvels2017}
{Deheuvels}, S., {Ouazzani}, R.~M., \& {Basu}, S. 2017, \aap, 605, A75

\bibitem[{{Di Mauro} {et~al.}(2016){Di Mauro}, {Ventura}, {Cardini}, {Stello},
  {Christensen-Dalsgaard}, {Dziembowski}, {Patern{\`o}}, {Beck}, {Bloemen},
  {Davies}, {De Smedt}, {Elsworth}, {Garc{\'{\i}}a}, {Hekker}, {Mosser}, \&
  {Tkachenko}}]{dimauro2016}
{Di Mauro}, M.~P., {Ventura}, R., {Cardini}, D., {et~al.} 2016, The
  Astrophysical Journal, 817, 65

\bibitem[{{Di Mauro} {et~al.}(2018){Di Mauro}, {Ventura}, {Corsaro}, \&
  {Lustosa De Moura}}]{dimauro2018}
{Di Mauro}, M.~P., {Ventura}, R., {Corsaro}, E., \& {Lustosa De Moura}, B.
  2018, \apj, 862, 9

\bibitem[{{Dyson} \& {Schutz}(1979)}]{dyson1979}
{Dyson}, J. \& {Schutz}, B.~F. 1979, Proceedings of the Royal Society of London
  Series A, 368, 389

\bibitem[{{Dziembowski} \& {Goode}(1992)}]{dziembowski1992}
{Dziembowski}, W.~A. \& {Goode}, P.~R. 1992, \apj, 394, 670

\bibitem[{{Eggenberger} {et~al.}(2019){Eggenberger}, {den Hartogh}, {Buldgen},
  {Meynet}, {Salmon}, \& {Deheuvels}}]{eggenberger2019}
{Eggenberger}, P., {den Hartogh}, J.~W., {Buldgen}, G., {et~al.} 2019, \aap,
  631, L6

\bibitem[{{Eggenberger} {et~al.}(2017){Eggenberger}, {Lagarde}, {Miglio},
  {Montalb{\'a}n}, {Ekstr{\"o}m}, {Georgy}, {Meynet}, {Salmon}, {Ceillier},
  {Garc{\'\i}a}, {Mathis}, {Deheuvels}, {Maeder}, {den Hartogh}, \&
  {Hirschi}}]{eggenberger2017}
{Eggenberger}, P., {Lagarde}, N., {Miglio}, A., {et~al.} 2017, \aap, 599, A18

\bibitem[{{Eggenberger} {et~al.}(2010){Eggenberger}, {Miglio}, {Montalban},
  {Moreira}, {Noels}, {Meynet}, \& {Maeder}}]{eggenberger2010}
{Eggenberger}, P., {Miglio}, A., {Montalban}, J., {et~al.} 2010, Astronomy and
  Astrophysics, 509, A72

\bibitem[{{Eggenberger} {et~al.}(2012){Eggenberger}, {Montalb{\'a}n}, \&
  {Miglio}}]{eggenberger2012}
{Eggenberger}, P., {Montalb{\'a}n}, J., \& {Miglio}, A. 2012, The Astrophysical
  Journal, 544, L4

\bibitem[{Fellay {et~al.}(2021)Fellay, Buldgen, Eggenberger, Khan, Salmon,
  Miglio, \& Montalbán}]{fellay2021}
Fellay, L., Buldgen, G., Eggenberger, P., {et~al.} 2021, \aap, 654, A133

\bibitem[{{Ferguson} {et~al.}(2005){Ferguson}, {Alexander}, {Allard}, {Barman},
  {Bodnarik}, {Hauschildt}, {Heffner-Wong}, \& {Tamanai}}]{ferguson2005}
{Ferguson}, J.~W., {Alexander}, D.~R., {Allard}, F., {et~al.} 2005, \apj, 623,
  585

\bibitem[{{Fuller} {et~al.}(2014){Fuller}, {Lecoanet}, {Cantiello}, \&
  {Brown}}]{fuller2014}
{Fuller}, J., {Lecoanet}, D., {Cantiello}, M., \& {Brown}, B. 2014, \apj, 796,
  17

\bibitem[{{Fuller} {et~al.}(2019){Fuller}, {Piro}, \& {Jermyn}}]{fuller2019}
{Fuller}, J., {Piro}, A.~L., \& {Jermyn}, A.~S. 2019, \mnras, 485, 3661

\bibitem[{{Gehan} {et~al.}(2018){Gehan}, {Mosser}, {Michel}, {Samadi}, \&
  {Kallinger}}]{gehan2018}
{Gehan}, C., {Mosser}, B., {Michel}, E., {Samadi}, R., \& {Kallinger}, T. 2018,
  \aap, 616, A24

\bibitem[{{Goupil} {et~al.}(2013){Goupil}, {Mosser}, {Marques}, {Ouazzani},
  {Belkacem}, {Lebreton}, \& {Samadi}}]{goupil2013}
{Goupil}, M.~J., {Mosser}, B., {Marques}, J.~P., {et~al.} 2013, Astronomy and
  Astrophysics, 549, A75

\bibitem[{Harris {et~al.}(2020)Harris, Millman, van~der Walt, Gommers,
  Virtanen, Cournapeau, Wieser, Taylor, Berg, Smith, Kern, Picus, Hoyer, van
  Kerkwijk, Brett, Haldane, Fernández~del Río, Wiebe, Peterson,
  Gérard-Marchant, Sheppard, Reddy, Weckesser, Abbasi, Gohlke, \&
  Oliphant}]{2020NumPy-Array}
Harris, C.~R., Millman, K.~J., van~der Walt, S.~J., {et~al.} 2020, Nature, 585,
  357–362

\bibitem[{{Hunter}(2007)}]{4160265}
{Hunter}, J.~D. 2007, Computing in Science Engineering, 9, 90

\bibitem[{{Iglesias} \& {Rogers}(1996)}]{iglesias1996}
{Iglesias}, C.~A. \& {Rogers}, F.~J. 1996, \apj, 464, 943

\bibitem[{{Jermyn} {et~al.}(2023){Jermyn}, {Bauer}, {Schwab}, {Farmer}, {Ball},
  {Bellinger}, {Dotter}, {Joyce}, {Marchant}, {Mombarg}, {Wolf}, {Sunny Wong},
  {Cinquegrana}, {Farrell}, {Smolec}, {Thoul}, {Cantiello}, {Herwig}, {Toloza},
  {Bildsten}, {Townsend}, \& {Timmes}}]{jermyn2023}
{Jermyn}, A.~S., {Bauer}, E.~B., {Schwab}, J., {et~al.} 2023, \apjs, 265, 15

\bibitem[{{K{\"a}pyl{\"a}}(2024)}]{kapyla2024}
{K{\"a}pyl{\"a}}, P.~J. 2024, \aap, 683, A221

\bibitem[{{Kjeldsen} \& {Bedding}(1995)}]{kjeldsen1995}
{Kjeldsen}, H. \& {Bedding}, T.~R. 1995, Astronomy and Astrophysics, 293, 87

\bibitem[{{Klion} \& {Quataert}(2017)}]{klion2017}
{Klion}, H. \& {Quataert}, E. 2017, \mnras, 464, L16

\bibitem[{{Li} {et~al.}(2024){Li}, {Deheuvels}, \& {Ballot}}]{li2024}
{Li}, G., {Deheuvels}, S., \& {Ballot}, J. 2024, \aap, 688, A184

\bibitem[{{Li} {et~al.}(2022){Li}, {Deheuvels}, {Ballot}, \&
  {Ligni{\`e}res}}]{li2022}
{Li}, G., {Deheuvels}, S., {Ballot}, J., \& {Ligni{\`e}res}, F. 2022, \nat,
  610, 43

\bibitem[{{Loi}(2021)}]{loi2021}
{Loi}, S.~T. 2021, \mnras, 504, 3711

\bibitem[{{Lynden-Bell} \& {Ostriker}(1967)}]{lynden1967}
{Lynden-Bell}, D. \& {Ostriker}, J.~P. 1967, \mnras, 136, 293

\bibitem[{{Maeder}(2009)}]{maeder2009}
{Maeder}, A. 2009, {Physics, Formation and Evolution of Rotating Stars}
  (Springer)

\bibitem[{{Marques} {et~al.}(2013){Marques}, {Goupil}, {Lebreton}, {Talon},
  {Palacios}, {Belkacem}, {Ouazzani}, {Mosser}, {Moya}, {Morel}, {Pichon},
  {Mathis}, {Zahn}, {Turck-Chi{\`e}ze}, \& {Nghiem}}]{marques2013}
{Marques}, J.~P., {Goupil}, M.~J., {Lebreton}, Y., {et~al.} 2013, \aap, 549,
  A74

\bibitem[{{Mosser} {et~al.}(2012){Mosser}, {Goupil}, {Belkacem}, {Marques},
  {Beck}, {Bloemen}, {De Ridder}, {Barban}, {Deheuvels}, {Elsworth}, {Hekker},
  {Kallinger}, {Ouazzani}, {Pinsonneault}, {Samadi}, {Stello}, {Garc{\'{\i}}a},
  {Klaus}, {Li}, {Mathur}, \& {Morris}}]{mosser2012}
{Mosser}, B., {Goupil}, M.~J., {Belkacem}, K., {et~al.} 2012, Astronomy and
  Astrophysics, 548, A10

\bibitem[{{Ong}(2024)}]{ong2024}
{Ong}, J.~M.~J. 2024, \apj, 960, 2

\bibitem[{{Ong} \& {Basu}(2020)}]{ong2020}
{Ong}, J.~M.~J. \& {Basu}, S. 2020, \apj, 898, 127

\bibitem[{{Ong} {et~al.}(2021){Ong}, {Basu}, \& {Roxburgh}}]{ong2021}
{Ong}, J.~M.~J., {Basu}, S., \& {Roxburgh}, I.~W. 2021, \apj, 920, 8

\bibitem[{{Ong} {et~al.}(2022){Ong}, {Bugnet}, \& {Basu}}]{ong2022}
{Ong}, J.~M.~J., {Bugnet}, L., \& {Basu}, S. 2022, \apj, 940, 18

\bibitem[{{Ouazzani} \& {Goupil}(2012)}]{ouazzani2012}
{Ouazzani}, R.~M. \& {Goupil}, M.~J. 2012, \aap, 542, A99

\bibitem[{{Paxton} {et~al.}(2011){Paxton}, {Bildsten}, {Dotter}, {Herwig},
  {Lesaffre}, \& {Timmes}}]{paxton2011}
{Paxton}, B., {Bildsten}, L., {Dotter}, A., {et~al.} 2011, \apjs, 192, 3,
  available at \url{http://mesa.sourceforge.net/}

\bibitem[{{Paxton} {et~al.}(2013){Paxton}, {Cantiello}, {Arras}, {Bildsten},
  {Brown}, {Dotter}, {Mankovich}, {Montgomery}, {Stello}, {Timmes}, \&
  {Townsend}}]{paxton2013}
{Paxton}, B., {Cantiello}, M., {Arras}, P., {et~al.} 2013, \apjs, 208, 4

\bibitem[{{Paxton} {et~al.}(2015){Paxton}, {Marchant}, {Schwab}, {Bauer},
  {Bildsten}, {Cantiello}, {Dessart}, {Farmer}, {Hu}, {Langer}, {Townsend},
  {Townsley}, \& {Timmes}}]{paxton2015}
{Paxton}, B., {Marchant}, P., {Schwab}, J., {et~al.} 2015, \apjs, 220, 15

\bibitem[{{Paxton} {et~al.}(2018){Paxton}, {Schwab}, {Bauer}, {Bildsten},
  {Blinnikov}, {Duffell}, {Farmer}, {Goldberg}, {Marchant}, {Sorokina},
  {Thoul}, {Townsend}, \& {Timmes}}]{paxton2018}
{Paxton}, B., {Schwab}, J., {Bauer}, E.~B., {et~al.} 2018, \apjs, 234, 34

\bibitem[{{Paxton} {et~al.}(2019){Paxton}, {Smolec}, {Schwab}, {Gautschy},
  {Bildsten}, {Cantiello}, {Dotter}, {Farmer}, {Goldberg}, {Jermyn}, {Kanbur},
  {Marchant}, {Thoul}, {Townsend}, {Wolf}, {Zhang}, \& {Timmes}}]{paxton2019}
{Paxton}, B., {Smolec}, R., {Schwab}, J., {et~al.} 2019, The Astrophysical
  Journal Supplement Series, 243, 10

\bibitem[{{Pijpers} \& {Thompson}(1992)}]{pijpers1992}
{Pijpers}, F.~P. \& {Thompson}, M.~J. 1992, \aap, 262, L33

\bibitem[{{Pijpers} \& {Thompson}(1994)}]{pijpers1994}
{Pijpers}, F.~P. \& {Thompson}, M.~J. 1994, \aap, 281, 231

\bibitem[{{Pin{\c{c}}on} {et~al.}(2017){Pin{\c{c}}on}, {Belkacem}, {Goupil}, \&
  {Marques}}]{pincon2017}
{Pin{\c{c}}on}, C., {Belkacem}, K., {Goupil}, M.~J., \& {Marques}, J.~P. 2017,
  \aap, 605, A31

\bibitem[{{Smeyers} \& {van Hoolst}(2010)}]{SmeyersVanHoolst}
{Smeyers}, P. \& {van Hoolst}, T. 2010, {Linear Isentropic Oscillations of
  Stars: Theoretical Foundations}, Vol. 371 (Springer Berlin, Heidelberg)

\bibitem[{{Spada} {et~al.}(2016){Spada}, {Gellert}, {Arlt}, \&
  {Deheuvels}}]{spada2016}
{Spada}, F., {Gellert}, M., {Arlt}, R., \& {Deheuvels}, S. 2016, \aap, 589, A23

\bibitem[{{Spruit}(2002)}]{spruit2002}
{Spruit}, H.~C. 2002, \aap, 381, 923

\bibitem[{{Su{\'a}rez} {et~al.}(2006){Su{\'a}rez}, {Goupil}, \&
  {Morel}}]{suarez2006}
{Su{\'a}rez}, J.~C., {Goupil}, M.~J., \& {Morel}, P. 2006, \aap, 449, 673

\bibitem[{{Takahashi} \& {Langer}(2021)}]{takahashi2021}
{Takahashi}, K. \& {Langer}, N. 2021, \aap, 646, A19

\bibitem[{{Theme{\ss}l} {et~al.}(2018){Theme{\ss}l}, {Hekker}, {Southworth},
  {Beck}, {Pavlovski}, {Tkachenko}, {Angelou}, {Ball}, {Barban}, {Corsaro},
  {Elsworth}, {Handberg}, \& {Kallinger}}]{themessl2018}
{Theme{\ss}l}, N., {Hekker}, S., {Southworth}, J., {et~al.} 2018, \mnras, 478,
  4669

\bibitem[{{Tisseur} \& {Meerbergen}(2001)}]{tisseur2001}
{Tisseur}, F. \& {Meerbergen}, K. 2001, SIAM Review, 43, 235

\bibitem[{{Townsend} {et~al.}(2018){Townsend}, {Goldstein}, \&
  {Zweibel}}]{townsend2018}
{Townsend}, R.~H.~D., {Goldstein}, J., \& {Zweibel}, E.~G. 2018, \mnras, 475,
  879

\bibitem[{{Townsend} \& {Teitler}(2013)}]{townsend2013}
{Townsend}, R.~H.~D. \& {Teitler}, S.~A. 2013, \mnras, 435, 3406

\bibitem[{{Triana} {et~al.}(2017){Triana}, {Corsaro}, {De Ridder}, {Bonanno},
  {P{\'e}rez Hern{\'a}ndez}, \& {Garc{\'{\i}}a}}]{triana2017}
{Triana}, S.~A., {Corsaro}, E., {De Ridder}, J., {et~al.} 2017, \aap, 602, A62

\bibitem[{{Unno} {et~al.}(1989){Unno}, {Osaki}, {Ando}, {Saio}, \&
  {Shibahashi}}]{unno}
{Unno}, W., {Osaki}, Y., {Ando}, H., {Saio}, H., \& {Shibahashi}, H. 1989,
  {Nonradial oscillations of stars}

\bibitem[{Virtanen {et~al.}(2020)Virtanen, Gommers, Oliphant, Haberland, Reddy,
  Cournapeau, Burovski, Peterson, Weckesser, Bright, {van der Walt}, Brett,
  Wilson, Millman, Mayorov, Nelson, Jones, Kern, Larson, Carey, Polat, Feng,
  Moore, {VanderPlas}, Laxalde, Perktold, Cimrman, Henriksen, Quintero, Harris,
  Archibald, Ribeiro, Pedregosa, {van Mulbregt}, \& {SciPy 1.0
  Contributors}}]{2020SciPy-NMeth}
Virtanen, P., Gommers, R., Oliphant, T.~E., {et~al.} 2020, Nature Methods, 17,
  261

\bibitem[{{W}es {M}c{K}inney(2010)}]{mckinney-proc-scipy-2010}
{W}es {M}c{K}inney. 2010, in {P}roceedings of the 9th {P}ython in {S}cience
  {C}onference, ed. {S}t\'efan van~der {W}alt \& {J}arrod {M}illman, 56 -- 61

\end{thebibliography}

	%-------------------------------------------------------------------
	%  Appendix
	%-------------------------------------------------------------------
	\begin{appendix}
		%--------------------------------------------------------------------
		
		%%%%%%%%%%%%%%%%%%%%%%%%%%%%%%%%%%%%%%%%%%%%%%%%
		\section{Perturbed equation of motion}\label{secperteqmo}
		
		A generic equation of motion of a rotating fluid in the inertial frame is 
		\begin{equation}\label{eqmotiongeneric}
			\pdv{\boldsymbol{v}}{t} + \left(\boldsymbol{v}\boldsymbol{\cdot}\nabla\right)\boldsymbol{v} = - \dfrac{\nabla p}{\rho} - \nabla \Phi\,,
		\end{equation}
		with $\boldsymbol{v} = \boldsymbol{\Omega} \times \boldsymbol{r}$.
		By introducing Eulerian perturbations of the quantities in Eq.~\eqref{eqmotiongeneric}, one obtains that $\boldsymbol{v}_0 = \boldsymbol{\Omega} \times \boldsymbol{r}$ and
		\begin{subequations}
			\begin{align}
				\left(\boldsymbol{v}_0\boldsymbol{\cdot}\nabla\right)\boldsymbol{v}_0 &= -\dfrac{\nabla p_0}{\rho_0} - \nabla \Phi_0\,,\label{eqgenequilib}\\
				\left(\pdv{\boldsymbol{v}}{t}\right)' + \left(\left(\boldsymbol{v}\boldsymbol{\cdot}\nabla\right)\boldsymbol{v}\right)' &= -\left(\dfrac{\nabla p}{\rho}\right)' - (\nabla \Phi)'\,,\label{eqpertgenmo}
			\end{align}
		\end{subequations}
		if one assumes that $\pdv{\boldsymbol{v}_0}{t} = 0$.
		It can be shown that
		\begin{subequations}
			\begin{align}
				\rho' &= - \nabla \boldsymbol{\cdot}\left(\rho_0\,\boldsymbol{\xi}\right)\,,\label{eqeulrho}\\
				p' &= -\Gamma_1\,p_0\,\left(\nabla\boldsymbol{\cdot}\boldsymbol{\xi}\right) - \boldsymbol{\xi}\boldsymbol{\cdot}\nabla p_0\,,\label{eqeulp}\\
				\boldsymbol{v}' &= \pdv{\boldsymbol{\xi}}{t} - \left(\boldsymbol{\xi}\boldsymbol{\cdot}\nabla\right)\boldsymbol{v}_0 + \left(\boldsymbol{v}_0\boldsymbol{\cdot}\nabla\right) \boldsymbol{\xi}\,,\label{eqeulv}
			\end{align}
		\end{subequations}
		hold to first order in $\boldsymbol{\xi}$ for adiabatic oscillations, where $\Gamma_1$ is the adiabatic index $(\partial \ln p / \partial \ln \rho)_{\rm S}$ (e.g. \citealt{SmeyersVanHoolst}).
		Substituting Eq.~\eqref{eqeulv} in Eq.~\eqref{eqpertgenmo}, one obtains
		\begin{equation}\label{intermediateeqmo}
			\begin{aligned}
				\pdv[2]{\boldsymbol{\xi}}{t} + \mathcal{R}_{v}\left(\pdv{\boldsymbol{\xi}}{t}\right) + \mathcal{V}_v\left(\boldsymbol{\xi}\right) = \left(\dfrac{\nabla p_0}{\rho_0^2}\right)\rho' -\dfrac{\nabla p'}{\rho_0} - \nabla \Phi'\,,
			\end{aligned}
		\end{equation}
		where $\mathcal{R}_v \equiv 2\left(\boldsymbol{v}_0\boldsymbol{\cdot}\nabla\right)$ and $\mathcal{V}_v\left(\boldsymbol{\xi}\right) \equiv \left(\boldsymbol{v}_0\boldsymbol{\cdot}\nabla\right)^2\boldsymbol{\xi} - \left(\boldsymbol{\xi}\boldsymbol{\cdot}\nabla\right)\left(\boldsymbol{v}_0\boldsymbol{\cdot}\nabla\right)\boldsymbol{v}_0$.
		Equation~\eqref{intermediateeqmo} is equal to Eq.~(1.1) of \citet{dyson1979} after multiplying by $\rho_0$ and substituting the relations~\eqref{eqgenequilib},~\eqref{eqeulrho} and~\eqref{eqeulp}.
		Dropping the second-order term $\mathcal{V}_v\left(\boldsymbol{\xi}\right)$ in Eq.~\eqref{intermediateeqmo}, and using the Ansatz $\boldsymbol{\xi} \propto \exp(-\di\,\omega\,t)$, one recovers Eq.~\eqref{eqosc}, because (see e.g. \citealt{aerts2010})
		\begin{equation}
			\mathcal{R}_{v}\left(\pdv{\boldsymbol{\xi}}{t}\right) \equiv -2\,\di\,\omega\left(\boldsymbol{v}_0\boldsymbol{\cdot}\nabla\right)\boldsymbol{\xi} = \omega\,\mathcal{R}\,\boldsymbol{\xi}\,.
		\end{equation}
		
		%%%%%%%%%%%%%%%%%%%%%%%%%%%%%%%%%%%%%%%%%%%%%%%%
		\section{Lowest-order perturbation theory}\label{secperttheo}
		
		The frequency perturbation of mode $i$ introduced by the rotational interaction with mode $j$ can be computed from evaluating the generic matrix element $\langle\boldsymbol{\xi}_{0,i}\,|\,\mathcal{R}\boldsymbol{\xi}_{0,j}\rangle$ defined in Sect.~\ref{secasy}:
		\begin{align}
			\delta\omega_{ij} &= \frac{1}{2}\langle\boldsymbol{\xi}_{0,i}\,|\,\mathcal{R}\,\boldsymbol{\xi}_{0,j}\rangle\nonumber\,,\\
			&=-\di\,\langle\boldsymbol{\xi}_{0,i}\,|\,(\di m\Omega+\boldsymbol{\Omega}\times)\,\boldsymbol{\xi}_{0,j}\rangle\,,\label{eqmatelement}
		\end{align}
		where the diagonal elements (i.e. $\delta\omega_{ii}$) signify the first order frequency perturbations in the inertial frame
		(e.g. Eq.~(19.48) of \citealt{unno}, Eq.~(3.261) and Eq.~(3.265) of \citealt{aerts2010}; see also \citetalias{dziembowski1992}, \citealt{deheuvels2017}, Sect. B.6.2 in \citealt{maeder2009}).
		Off-diagonal elements measure contributions from NDE \citepalias{dziembowski1992}.
		We define the unnormalised displacement vector as
		\begin{align}
			\boldsymbol{\tilde\xi} = \left[\tilde\xi_\rr Y_\ell^m\boldsymbol{e}_r+\tilde\xi_\hh\left(\frac{\partial Y_\ell^m}{\partial \theta}\boldsymbol{e}_\theta+\frac{1}{\sin\theta}\frac{\partial Y_\ell^m}{\partial \phi}\boldsymbol{e}_\phi\right)\right]{\rm e}^{-\di\omega t}\,.
		\end{align}
		Here, $Y_\ell^m \equiv Y_\ell^m(\theta,\phi)$ denote the spherical harmonics as defined in \cite{aerts2010}  with $\mathbf{e}_r, \mathbf{e}_\theta$ and $\mathbf{e}_\phi$ being the unit vectors of spherical coordinates and $\tilde\xi_\rr\equiv\tilde\xi_\rr(r),\,\tilde\xi_\hh\equiv\tilde\xi_\hh(r)$ denoting the unnormalised radial and horizontal displacement functions, respectively.
		Now we can compute the normalisation of the eigenfunction \citep[Eq.~(3.347) of][]{aerts2010}:
		\begin{align}
			&|\boldsymbol{\tilde\xi}_i|^2=\int_0^R\left[\tilde\xi_{\rr,i}^*\tilde\xi_{\rr,i}+\ell(\ell+1)\,\tilde\xi_{\hh,i}^*\tilde\xi_{\hh,i}\right]\rho_0(r)\,r^2{\rm d}r\,.
		\end{align}
		To compute the matrix element defined by Eq.~\eqref{eqmatelement}, we normalise the radius-dependent parts of the displacement vector components $\xi_\rr,\xi_\hh$ such that $|\boldsymbol{\xi}|=1$. For the sake of readability we drop the index $0$ for the unperturbed modes in the remainder of this section. By evaluating the cross product originating from the Coriolis force \citep[e.g. Eq.~(3.342) of][]{aerts2010} we find
		\begin{align}
			&&&\langle\boldsymbol{\xi}_i\,|\,(\di m\Omega+\boldsymbol{\Omega}\times)\boldsymbol{\xi}_j\rangle\nonumber\\
			&&=&\int_V\rho_0(r)\,\boldsymbol{\xi}_i^*\cdot\di m\Omega(r)\,\boldsymbol{\xi}_j\,{\rm d}V + \int_V\rho_0(r)\,\boldsymbol{\xi}_i^*\cdot\boldsymbol{\Omega}\times\boldsymbol{\xi}_j\,{\rm d}V\,, \nonumber\\
			&&=&\di m \int_V\rho_0(r)\,\boldsymbol{\xi}_i^*\cdot\Omega(r)\,\boldsymbol{\xi}_j{\rm d}V \nonumber\\
			&&&+\int_V\rho_0(r)\,\boldsymbol{\xi}_i^*\cdot\nonumber\\
			&&&\phantom{====}\left(\Omega(r)\left[-\xi_{\hh,j}\frac{\partial Y_\ell^m}{\partial \phi}\boldsymbol{e}_r-\xi_{\hh,j}\frac{\cos\theta}{\sin\theta}\frac{\partial Y_\ell^m}{\partial \phi}\boldsymbol{e}_\theta\right.\right.\nonumber\\
			&&&\phantom{====}\left.\left.+\left(\xi_{\rr,j}\sin\theta Y_\ell^m+\xi_{\hh,j}\cos\theta\frac{\partial Y_\ell^m}{\partial \theta}\right)\boldsymbol{e}_\phi\right]\right)\,{\rm d}V\,.
		\end{align}
		This thus becomes:
		\begin{align}
			&&&\langle\boldsymbol{\xi}_i\,|\,(\di m\Omega+\boldsymbol{\Omega}\times)\boldsymbol{\xi}_j\rangle\nonumber\\
			&&=&\di m \int_V\rho_0(r)\,\boldsymbol{\xi}_i^*\cdot\Omega(r)\,\boldsymbol{\xi}_j{\rm d}V&\nonumber\\
			&&&+\int_V\rho_0(r)\,\Omega(r)&\nonumber\\
			&&&\phantom{====}\left[\left(\xi_{\rr,i}Y_\ell^m\right)^*\left(-\xi_{\hh,j}\frac{\partial Y_\ell^m}{\partial\phi}\right)+\left(\xi_{\hh,i}\frac{\partial Y_\ell^m}{\partial\theta}\right)^*\left(-\xi_{\hh,j}\frac{\cos\theta}{\sin\theta}\frac{\partial Y_\ell^m}{\partial\phi}\right)\right.&\nonumber\\
			&&&\phantom{====}\left.+\left(\xi_{\hh,i}\frac{1}{\sin\theta}\frac{\partial Y_\ell^m}{\partial\phi}\right)^*\left(\xi_{\rr,j}\sin\theta Y_\ell^m+\xi_{\hh,j}\cos\theta\frac{\partial Y_\ell^m}{\partial\theta}\right)\right]\,{\rm d}V\,.&
		\end{align}
		In the second integral all terms contain the $\phi$ derivative of $Y_\ell^m$, resulting in a multiplicative factor $\di m$ for $Y_\ell^m$. Note that the sign changes for some of the derivative terms due to complex conjugation. The integral in $\phi$ results in a multiplicative factor of $2\pi$, as $Y_\ell^m$ and derivatives thereof always appear in pairs of complex conjugates. We can hence proceed to write:
		\begin{align}
			&&&\langle\boldsymbol{\xi}_i\,|\,(im\Omega+\boldsymbol{\Omega}\times)\boldsymbol{\xi}_j\rangle\nonumber\\
			&&=&2\pi c_{\ell m}^2\di m\int_0^R\Omega(r)\,\rho_0(r)\int_0^\pi\nonumber\\
			&&&\phantom{}\left[\xi_{\rr,i}^*\xi_{\rr,j}\left(P_\ell^m\right)^2+\xi_{\hh,i}^*\xi_{\hh,j}\phantom{^*}\left[\left(\frac{\partial P_\ell^m}{\partial\theta}\right)^2+\frac{m^2}{\sin\theta}\left(P_\ell^m\right)^2\right]\right]\,\sin\theta{\rm d}\theta \,r^2{\rm d}r\nonumber\\
			&&&+2\pi c_{\ell m}^2\di m\int_0^R\Omega(r)\,\rho_0(r)\int_0^\pi\nonumber\\
			&&&\phantom{}\left[-\left(\xi_{\rr,i}^*\xi_{\hh,j}+\xi_{\hh,i}^*\xi_{\hh,j}\right)\left(P_\ell^m\right)^2-2\xi_{\hh,i}^*\xi_{\hh,j}\frac{\cos\theta}{\sin\theta}\frac{\partial P_\ell^m}{\partial\theta}P_\ell^m\right]\,\sin\theta{\rm d}\theta \,r^2{\rm d}r\,,
		\end{align}
		where we use
		\begin{align}
			c_{\ell m}^2=\frac{(2l+1)(l-m)!}{4\pi(l+m)!}\,.
		\end{align}
		Evaluating the integrals over the co-latitude $\theta$, one finds:
		\begin{align}
			&\langle\boldsymbol{\xi}_i\,|\,(\di m\Omega+\boldsymbol{\Omega}\times)\boldsymbol{\xi}_j\rangle\nonumber\\
			=&\di m\int_0^R\Omega(r)\,\rho_0(r)\nonumber\\
			&\phantom{\di m}\left[\xi_{\rr,i}^*\xi_{\rr,j}+\ell(\ell+1)\,\xi_{\hh,i}^*\xi_{\hh,j}-\xi_{\hh,i}^*\xi_{\hh,j}-\left(\xi_{\rr,i}^*\xi_{\hh,j}+\xi_{\hh,i}^*\xi_{\rr,j}\right)\right]r^2{\rm d}r\,.\label{eqperturbation}
		\end{align}
		
		Together with Eq.~\eqref{eqperturbation} and the factor of $-\di$ from the operator (Eq.~\eqref{eqmatelement}) we can write down the expression for the perturbation due to rotation and define the kernel $\mathcal{K}_{ij}$:
		\begin{align}
			&\delta\omega_{ij}=m\int_0^R\mathcal{K}_{ij}(r)\,\Omega(r){\rm d}r\,,\nonumber\\
			&\phantom{\delta\omega_{ij}}=m\int_0^R\Omega(r)\,\rho_0(r)\nonumber\\
			&\phantom{\delta\omega_{ij}=m}\left[\xi_{\rr,i}^*\xi_{\rr,j}+\left(L^2-1\right)\xi_{\hh,i}^*\xi_{\hh,j}-\left(\xi_{\rr,i}^*\xi_{\hh,j}+\xi_{\hh,i}^*\xi_{\rr,j}\right)\right]r^2{\rm d}r\,,\label{eqcrosskernel}
		\end{align}
		(e.g. \citealt{deheuvels2017,ong2022}) with $L^2=\ell(\ell+1)$. Given the normalisation of the eigenfunctions $\boldsymbol{\xi}_i$ and $\boldsymbol{\xi}_j$ the total integral of the kernels $\mathcal{K}_{ij}$ is written as
		\begin{align}
			\beta_{ij}&=\frac{\int_0^R\left[\tilde\xi_{\rr,i}^*\tilde\xi_{\rr,j}+\left(L^2-1\right)\tilde\xi_{\hh,i}^*\tilde\xi_{\hh,j}-\left(\tilde\xi_{\rr,i}^*\tilde\xi_{\hh,j}+\tilde\xi_{\hh,i}^*\tilde\xi_{\rr,j}\right)\right]\rho_0(r)\,r^2{\rm d}r}{|\boldsymbol{\tilde\xi}_i||\boldsymbol{\tilde\xi}_j|}\,.\label{eqbetaij}
		\end{align}
		The `standard' rotational kernel (see Eq.~\eqref{eqsplitting}) is obtained from that general expression for $i=j$:
		\begin{align}
			\delta\omega_{ii}
			&=m\int_0^R\mathcal{K}_{ii}(r)\,\Omega(r){\rm d}r\,,\nonumber\\
			&= \frac{m\int_0^R\left[\tilde\xi_{\rr,i}^2+\left(L^2-1\right)\tilde\xi_{\hh,i}^2-2\tilde\xi_{\rr,i}\tilde\xi_{\hh,i}\right]\rho_0(r)\,\Omega(r)\,r^2{\rm d}r}{\int_0^R\left[\tilde\xi_{\rr,i}^2+\ell(\ell+1)\,\tilde\xi_{\hh,i}^2\right]\rho_0(r)\,r^2{\rm d}r}\,,
		\end{align}
		assuming that the radial and horizontal displacement functions are real.
		
		%%%%%%%%%%%%%%%%%%%%%%%%%%%%%%%%%%%%%%%%%%%%%%%%
		\section{Solution of the NDE QEP}\label{secqep}
		We solve the QEP of Eq.~\eqref{eqoscmat} by transforming it into a generalised linear eigenvalue problem of the form \citep[this is the first companion form of][with their N equal to $-\mathbf{I}$]{tisseur2001}:
		\begin{align}
			\mathbf{A}\mathbf{a}'=\omega\mathbf{B}\mathbf{a}'\,,\label{eqgep}
		\end{align}
		where we define new (right) eigenvectors\footnotemark[2] $\mathbf{a}'$
		\footnotetext[2]{As the matrices $\mathbf{L}$, $\mathbf{R}$ and $\mathbf{D}$ are real symmetric, the sets of left and right QEP eigenvectors coincide \citep[e.g.][]{tisseur2001}.} as
		\begin{align}
			\mathbf{a}'=
			\begin{pmatrix}
				\mathbf{a}\\
				\omega\mathbf{a}
			\end{pmatrix}\,.
		\end{align}
		The block matrices $\mathbf{A}$ and $\mathbf{B}$ are defined as
		\begin{subequations}
			\begin{align}
				\mathbf{A} &= \begin{pmatrix}
					\mathbf{0} & -\mathbf{I}\\
					-\mathbf{L}&-\mathbf{R}
				\end{pmatrix}\,,\\
				\mathbf{B}&= \begin{pmatrix}
					-\mathbf{I} & \mathbf{0}\\
					\mathbf{0}&-\mathbf{D}
				\end{pmatrix}\,.
			\end{align}
		\end{subequations}
		By definition, $\mathbf{D} = \mathbf{I}$ (see Eq.~\eqref{eqdij}) in this QEP.
		Hence, the generalised eigenvalue problem posed in Eq.~\eqref{eqgep} can be written as the augmented linear eigenvalue problem
		\begin{equation}
			\mathbf{A}\mathbf{a}'=-\omega\mathbf{a}'\,.\label{eqaugep}
		\end{equation}
		
		When solving the eigenvalue problem numerically using methods implemented in {\sc SciPy}, we find eigenvalues $\omega$ and $-\omega$ that are associated with the eigenfunction $\mathbf{\xi}$ defined by Eq.~\eqref{eqeigenfunction}.
		The perturbed mode frequencies (when we include NDE) are then equal to the eigenvalues $\omega$ of Eq.~\eqref{eqaugep}, and the eigenvectors $\mathbf{a}$ can be extracted from the augmented eigenvectors $\mathbf{a}'$.
		%%%%%%%%%%%%%%%%%%%%%%%%%%%%%%%%%%%%%%%%%%%%%%%%
		\section{Generalised NDE treatment of \citetalias{dziembowski1992}}\label{sectilde}
		
		Instead of solving the QEP posed in Eq.~\eqref{eqoscmat} by linearising it into a generalised eigenvalue problem $\mathbf A\mathbf{a}'=\omega \mathbf B\mathbf{a}'$, one can also generalise the treatment of NDE in Section 3.3 of \citetalias{dziembowski1992}.
		We define the unperturbed frequency difference
		\begin{equation}\label{eqfreqdiff}
			\Delta\omega_i = \omega_{0,i} - \omega_{0,1}\,,
		\end{equation}
		and the perturbed frequency as
		\begin{equation}\label{eqtildedef}
			\omega = \omega_{0,1} + \Tilde{\omega}\,,
		\end{equation}
		where $\omega_{0,i}$ denotes the $i$-th unperturbed mode frequency of the near-degenerate modes and $\Tilde{\omega}$ is a perturbation to the frequency analogous to $\Tilde{\omega}_1$ of \citetalias{dziembowski1992}.
		If $\omega \neq 0$, we can equivalently write Eq.~\eqref{eqoscmat} as
		\begin{equation}
			\omega\left(\dfrac{\mathbf L}{\omega}+\mathbf {R}-\omega\mathbf{D}\right)\mathbf{a}\equiv \omega\mathbf{Z}\mathbf{a}=0=\mathbf{Z}\mathbf{a}\,,\label{eqoscmattilde}
		\end{equation}
		so that the elements on the diagonal of $\mathbf{Z}$ are
		\begin{equation}\label{eqdiagZ}
			Z_{ii} = \Delta\omega_i -\omega_{0,i} - \Tilde{\omega} + 2\delta\omega_{ii} + \dfrac{\omega_{0,i}}{\left(1 + \frac{[\Tilde{\omega}-\Delta\omega_i]}{\omega_{0,i}}\right)}\,,
		\end{equation}
		where we have used Eqs.~\eqref{eqfreqdiff} and \eqref{eqtildedef}.
		For nearly degenerate frequencies, $\Tilde{\omega}$ and $\Delta\omega_i$ are small compared to $\omega_{0,i}$, so that the diagonal elements of $\mathbf{Z}$ can be approximated by
		\begin{equation}\label{eqdiagZapprox}
			Z_{ii} \approx 2\left(\Delta\omega_i - \Tilde{\omega} + \delta\omega_{ii}\right)\,.
		\end{equation}
		Denoting this approximation of $\mathbf{Z}$ as $\mathbf{Z}_1$, we can approximate the solution of the QEP posed in Eq.~\eqref{eqoscmat} by
		\begin{equation}
			\left[\dfrac{1}{2}\mathbf{Z}_1 + \text{diag}\left(\Tilde{\omega}\right)\right]\mathbf{a} \equiv \mathbf{Z}_2 \, \mathbf{a} = \Tilde{\omega}\, \mathbf{a}\,.
		\end{equation}
		Solving the eigenvalue problem $\mathbf{Z}_2 \, \mathbf{a} = 0$ thus yields the perturbations $\Tilde{\omega}$ as eigenvalues, and $\mathbf{a}$ as eigenvectors.
		%%%%%%%%%%%%%%%%%%%%%%%%%%%%%%%%%%%%%%%%%%%%%%%%%%%%%%%%%%%%%%%%%%%%%%%%%%%%%
		\section{Comparison to the $\pi$-$\gamma$ decomposition}\label{secpigamma}
		\begin{figure*}
			\centering
			\includegraphics[]{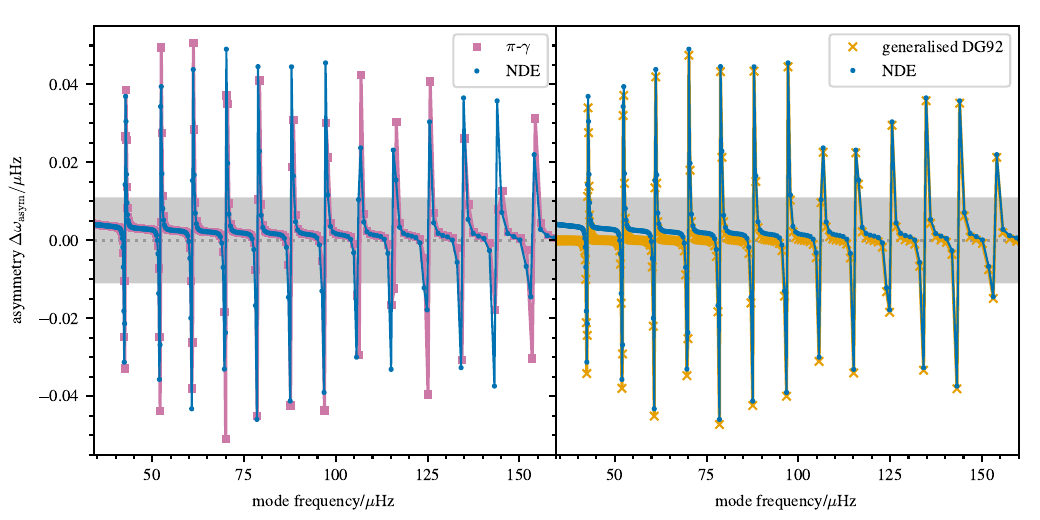}
			\caption{Comparison of rotational splitting asymmetries $\Delta\omega_{\rm asym}$ computed from the NDE, generalised \citetalias{dziembowski1992} and $\pi$-$\gamma$ formalisms. Results for the $\pi$-$\gamma$ formalisms and the generalised \citetalias{dziembowski1992} are shown in the left and right panel, respectively.  Here, the evolved model was used to compute the synthetic data. The shaded band indicates two times the minimum uncertainty of the averaged splittings obtained from Eq.~\eqref{eqsplittingsymm}. The multiplication by two is necessary because $\Delta\omega_{\rm asym}$ is twice the frequency difference between the averaged and actual rotational splittings. As a minimum uncertainty of the observed frequencies we assume the frequency resolution. Then the minimum uncertainty of the averaged splittings is computed from error propagation as $\sigma_{\delta\omega, {\rm min}}/(2\pi)=1/\sqrt{2}\,\delta\nu_{\rm res}=1/\sqrt{2}\times0.0078~\mu$Hz, given the four years of Kepler observations.}
			\label{figasymmetrycomparison}
		\end{figure*}
		To validate the results that we obtained from the NDE formalism discussed in Sect.~\ref{secasy}, we computed oscillation frequencies following the $\pi$-$\gamma$ decomposition and recoupling described by \cite{ong2020,ong2021}. To include the effects of rotation we solve Eq.~\eqref{eqoscmat} in the $\pi$-$\gamma$ basis. Here, the matrices are defined as follows:
		\begin{subequations}
			\begin{align}
				\mathbf{L}&=\begin{pmatrix}
					\mathbf{\Omega}_\pi^2-\mathbf{R}_{\pi\pi}&\mathbf{\Omega}_\pi^2\mathbf{D}_{\pi\gamma}-\mathbf{R}_{\pi\gamma}\\
					(\mathbf{\Omega}_\pi^2\mathbf{D}_{\pi\gamma}-\mathbf{R}_{\pi\gamma})^T&\mathbf{\Omega}_\gamma^2-\mathbf{R}_{\gamma\gamma}
				\end{pmatrix}\,,\label{eqlpigamma}\\
				\mathbf{R} &=\begin{pmatrix}
					2\boldsymbol{\delta\omega_\pi}&\boldsymbol{0}\\
					\boldsymbol{0}&2\boldsymbol{\delta\omega_\gamma}
				\end{pmatrix}\,,\label{eqrpigamma}\\
				\mathbf{D}&=\begin{pmatrix}
					\boldsymbol{I}&\mathbf{D}_{\pi\gamma}\\
					\mathbf{D}_{\pi\gamma}^T&\boldsymbol{I}
				\end{pmatrix}\,,\label{eqdpigamma}
			\end{align}
		\end{subequations}
		where $\boldsymbol{\Omega_\pi}^2=\text{diag}(\omega_{\pi,i}^2)$, $\boldsymbol{\Omega_\gamma}^2=\text{diag}(\omega_{\gamma,i}^2)$ are diagonal matrices containing the unperturbed eigenfrequencies of the $\pi$ and $\gamma$ modes and $\boldsymbol{\delta\omega_\pi}=\text{diag}(\delta\omega_{\pi,i})$ and $\boldsymbol{\delta\omega_\gamma}=\text{diag}(\delta\omega_{\gamma,i})$ are diagonal matrices containing the rotational splittings of the $\pi$ and $\gamma$ modes computed according to Eq.~\eqref{eqsplitting}. For the definitions of the other matrix elements $\mathbf{R}_{\pi\pi},\mathbf{R}_{\gamma\gamma},\mathbf{R}_{\pi\gamma},\mathbf{D}_{\pi\gamma}$ we refer to \cite{ong2021}. We note that we have switched the sign in our definition of $\mathbf{L}$ in comparison to \cite{ong2021} in order to comply with the definition of Eq.~\eqref{eqoscmat}.
		We also take the time dependence of the oscillations ($\propto\exp(-\di\omega t)$) into account. To solve the QEP posed by Eq.~\eqref{eqoscmat} and Eqs.~\eqref{eqlpigamma}, ~\eqref{eqrpigamma} and \eqref{eqdpigamma} we transform it into a generalised Eigenvalue problem as in Eq.~\eqref{eqgep}. In contrast to the NDE QEP defined in Appendix~\ref{secqep} we cannot rewrite the $\pi$-$\gamma$ QEP into an augmented linear Eigenvalue problem.
		
		Using the frequencies and eigenfunctions of the $\pi$- and $\gamma$-modes we computed the mixed mode frequencies for each azimuthal order $m$ separately. Subsequently, we computed rotational splittings for each $m$, according to Eqs.~\eqref{eqsplittingsasym1} and \eqref{eqsplittingsasym2} and the associated asymmetry $\Delta\omega_{\rm asym}$ as the difference between the rotational splittings of different $m$ according to Eq.~\eqref{eqasymmetry}. In Fig.~\ref{figasymmetrycomparison} we compare the asymmetries computed from the $\pi$-$\gamma$ and NDE formalisms. We find good agreement both in terms of magnitude and find similar patterns of the asymmetries as a function of mode frequency. The formalisms compared here are independent methods to compute the rotational splittings. We consider this agreement an important validation of these methods.

	\end{appendix}
	
\end{document}